\documentclass[conference,compsoc]{IEEEtran}

\ifCLASSOPTIONcompsoc
  \usepackage[nocompress]{cite}
\else
  \usepackage{cite}
\fi

\usepackage{tabularx} 
\usepackage{arydshln}
\usepackage{pifont}
\usepackage[linesnumbered,ruled,vlined,boxed]{algorithm2e}
\usepackage{algpseudocode,float}
\usepackage{textcomp}
\usepackage{float}
\usepackage{amsfonts}
\usepackage{amssymb}
\usepackage{amsmath}
\usepackage{mathdots}
\usepackage{mathtools}
\usepackage{amsthm}
\usepackage{stfloats}
\usepackage{setspace}
\usepackage{bbm}
\usepackage{bm}
\usepackage{multirow}
\usepackage[numbers,sort&compress]{natbib}
\usepackage{mathrsfs}
\usepackage[table]{xcolor}
\usepackage{url}
\usepackage{enumitem}
\usepackage{multirow,tabularx}
\usepackage{array}
\usepackage{booktabs}
\usepackage{url}
\usepackage[switch]{lineno}
\usepackage{graphicx}
\usepackage{subfigure}
\usepackage{epstopdf}
\usepackage{color}
\usepackage{makecell}
\definecolor{dark_green}{RGB}{ 65,154, 35}
\newcommand{\sysname}{\textsc{Spa }}
\def\BibTeX{{\rm B\kern-.05em{\sc i\kern-.025em b}\kern-.08em
		T\kern-.1667em\lower.7ex\hbox{E}\kern-.125emX}}
\begin{document}

\title{\sysname: Towards More Stealth and Persistent Backdoor Attacks in Federated Learning\\

}

 \author{
 \IEEEauthorblockN{
 Chengcheng Zhu\IEEEauthorrefmark{1},
 Ye Li\IEEEauthorrefmark{2},
 Bosen Rao\IEEEauthorrefmark{3},
 Jiale Zhang\IEEEauthorrefmark{3},
 Yunlong Mao\IEEEauthorrefmark{1}, and 
 Sheng Zhong\IEEEauthorrefmark{1}
 }

 \IEEEauthorblockA{
 \IEEEauthorrefmark{1}State Key Laboratory for Novel Software Technology, Nanjing University, Nanjing 210023, China}

 \IEEEauthorblockA{\IEEEauthorrefmark{2}College of Computer Science and Technology, Nanjing University of Aeronautics and Astronautics, Nanjing, 211106, China}
 
 \IEEEauthorblockA{\IEEEauthorrefmark{3}School of Information Engineering, Yangzhou University, Yangzhou, 225127, China}

 }

\maketitle

\begin{abstract}
Federated Learning (FL) has emerged as a leading paradigm for privacy-preserving distributed machine learning, yet the distributed nature of FL introduces unique security challenges, notably the threat of backdoor attacks. Existing backdoor strategies predominantly rely on end-to-end label supervision, which, despite their efficacy, often results in detectable feature disentanglement and limited persistence. In this work, we propose a novel and stealthy backdoor attack framework, named \sysname, which fundamentally departs from traditional approaches by leveraging feature-space alignment rather than direct trigger-label association. Specifically, \sysname reduces representational distances between backdoor trigger features and target class features, enabling the global model to misclassify trigger-embedded inputs with high stealth and persistence. We further introduce an adaptive, adversarial trigger optimization mechanism, utilizing boundary-search in the feature space to enhance attack longevity and effectiveness, even against defensive FL scenarios and non-IID data distributions. Extensive experiments on various FL benchmarks demonstrate that \sysname consistently achieves high attack success rates with minimal impact on model utility, maintains robustness under challenging participation and data heterogeneity conditions, and exhibits persistent backdoor effects far exceeding those of conventional techniques. Our results call urgent attention to the evolving sophistication of backdoor threats in FL and emphasize the pressing need for advanced, feature-level defense techniques.
\end{abstract}


\section{Introduction}\label{sec:introduction}

Federated Learning \cite{FL_mcmahan_2017, HeteroFair_Li_2024}, a promising distributed learning paradigm, consists of a group of data-owning clients under the orchestration of a central server, which allows each client to collaboratively train a global model while keeping their data local. For each iteration in FL, clients locally train models based on local training data and then upload these models or gradients to the server, aggregating them to obtain a global model. The global model is then propagated back to the clients for the next training iteration. As such, FL has become an emerging privacy-preserving trend with many applications in popular mobile apps, such as Google’s GBoard \cite{GBoard_hard_2018}, Apple’s QuickType \cite{QuickType_Apple_2019}, and various data-sensitive application domains including financial \cite{Financial_long_2020}, medical \cite{Medical_sheller_2019}, and security \cite{security} scenarios.

The medicine can be a poison. Despite enhanced privacy, the distributed nature of FL systems makes supervising the underlying processes of the local training hard, leading to a series of new security issues \cite{feng2023infer1,fu2022infer2,ma2022posion1,fang2020posion2}. Among them, backdoor attacks \cite{A3FL_Zhang_2024, backdoorFL_bagdasaryan_2020, Neurotoxin_zhang_2022, Chameleon_dai_2023, Advdoor_zhang_2021, Darkfed_li_2024}, have gained significant attention due to their stealth and practical effectiveness, becoming one of the most threatening of the FL system. Specifically, an attacker injects a backdoor trigger into a subset of its training data and labels the backdoored data with an attacker-chosen target label. The adversary then submits the backdoored model trained on the poisoned data, causing the global model to inherit the backdoor function. As a result, the global model performs normally on clean test inputs but misclassifies any input stamped with the attacker-chosen backdoor trigger as the specified target class. 

\begin{figure}
	\centering
	\includegraphics[width=1\linewidth]{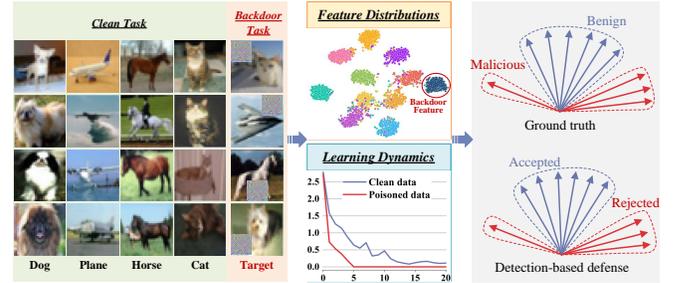}
	\caption{Anomalies caused by backdoor attacks in federated learning.}
	\label{F1}
	\vspace{-3mm}
\end{figure}

Previous works have successfully implanted backdoors in FL models. To ensure the effectiveness and stealth of the backdoor, attackers employ various strategies such as trigger optimization \cite{A3FL_Zhang_2024, Advdoor_zhang_2021, Darkfed_li_2024} and model constraints \cite{backdoorFL_bagdasaryan_2020, Neurotoxin_zhang_2022, Chameleon_dai_2023} to prevent the backdoor from being removed during federated aggregation and to bypass advanced backdoor defenses. However, fundamentally, these methods still rely on an end-to-end approach that establishes a strong connection between the trigger and the target label to inject the backdoor (as illustrated in Figure \ref{F1}). In this scenario, the backdoor task and the clean task are treated as two distinct tasks \cite{li2021anti}. Since the backdoor trigger is more apparent than normal features and frequently occurs, the backdoor task converges faster, and the backdoor features become highly disentangled from other features. These characteristics result in significant differences between models trained by malicious participants and those trained by benign participants, which form the basis for existing defense mechanisms \cite{FLAME_Nguyen_2022}. Although some studies attempt to bypass defenses by strictly regulating the difference between local and global model weights, this approach inherently limits the effectiveness of the backdoor. This line of exploration brings us to a fundamental yet often overlooked question, \textit{``Is it possible to inject backdoors efficiently and stealthily without relying on traditional end-to-end supervised training methods?''}

From the perspective of different feature learning paradigms, supervised learning leverages high-quality labeled data, updating model parameters in an end-to-end manner by minimizing the difference between predicted and true labels. This process allows the model to learn discriminative features for each class, resulting in distinct and separable clusters in the feature space. On the other hand, self-supervised learning approaches, such as contrastive learning \cite{khosla2020contrastive}, learn more generalized representations without labels by pulling similar features closer and pushing dissimilar features apart. Drawing inspiration from this feature alignment mechanism, instead of directly associating trigger features with the target label, \textit{there is a new way that establishes a connection between backdoor trigger features and target class features by minimizing their distance}, thereby tricking the model into recognizing the backdoor features as normal features of the target class.

In this paper, we propose a more stealthy and persistent backdoor attack method, named SPA, which comprises two key components: backdoor injection and backdoor enhancement. In the backdoor injection phase, we deviate from the traditional approach of establishing a strong association between triggers and target labels through end-to-end supervised training. Instead, we achieve feature alignment by reducing the feature distribution distance between trigger-embedding samples and target-class samples, ensuring that the backdoor model produces highly similar feature representations for them, leading the model to classify trigger-embedding inputs into the target class. In the backdoor enhancement phase, inspired by adversarial examples attacks, we search for noise features on the boundary of the target class distribution within the feature space of the global model. Such noise, which naturally shifts the model’s predictions toward the target class, is used as backdoor triggers to address the challenge of aligning fixed, out-of-distribution triggers with the target distribution, thereby enhancing the effectiveness and persistence of the backdoor.

For evaluations, we conduct extensive experiments to validate the effectiveness, stealthiness, and persistence of \sysname in FL environments. Our comprehensive experimental evaluation encompasses several critical dimensions of analysis: First, in comparison with state-of-the-art (SOTA) methods, \sysname demonstrates superior attack success rates while maintaining minimal impact on model utility. To confirm the practical applicability, we examine its effectiveness across varying non-independent and identically distributed (non-IID) scenarios, diverse model architectures, and multiple adversary participation scenarios. The results consistently show that  \sysname maintains high ASR even in challenging scenarios with highly skewed data distributions and limited adversary participation, showcasing its robustness to real-world FL conditions. Our persistence analysis reveals a particularly concerning property of our attack, demonstrating remarkable longevity even after the adversary ceases participation. When evaluating different attack timing strategies, we observe that  \sysname remains effective even after 1000 training rounds without adversary involvement, significantly outlasting conventional backdoor approaches that typically decay after 100-200 rounds.

In more complex attack scenarios, such as multi-label attacks and various trigger implementations,  \sysname maintains consistent performance across different trigger types and successfully targets multiple labels simultaneously without compromising the primary attack objectives. Through parameter sensitivity studies and ablation experiments, we systematically analyze the contribution of each component in our approach. Additionally, our analysis of different constraint norms demonstrates that the distance measures employed in  \sysname outperform alternative measures like $L_2$ norm and KL-divergence in balancing attack effectiveness and model utility. Collectively, these comprehensive experiments demonstrate that our backdoor attack methodology poses a significant security concern for federated learning systems, highlighting the urgent need for advanced defense mechanisms to counter such sophisticated threats.

The main contributions of this paper are summarized as follows:
\begin{itemize}
    \item \textbf{A Novel Backdoor Attack Paradigm.} We propose \sysname, a novel backdoor attack paradigm that abandons traditional end-to-end supervised training in favor of feature alignment between trigger patterns and target class distributions. This fundamental shift in attack methodology enables the backdoor to be more stealthy and persistent.

    \item \textbf{Adversarial Dynamic Trigger Optimization.} \sysname introduces an innovative adversarial trigger optimization mechanism, which enhances backdoor effectiveness by identifying natural adversarial noise patterns within the feature space to serve as triggers. Moreover, by leveraging feature consistency constraints, backdoor intensity can be dynamically adjusted using the latest global model, thereby enhancing attack stealth while maintaining effectiveness.

    \item \textbf{Comprehensive Empirical Validations.} We conduct extensive experiments validating the stealth and persistence characteristics of our proposed method. Our results confirm the applicability and effectiveness of \sysname across various federated learning configurations and backdoor settings, highlighting the urgent need for advanced defense mechanisms to counter such sophisticated threats.
    
\end{itemize}

The remaining of this paper is organized as follows. Section \ref{Sec: Related Work} introduces the related works of backdoor attacks and defenses. The problem formulation and proposed method are discussed in Section \ref{Sec: Problem} and \ref{Sec: Methodology}. Section \ref{Sec: Experimental Evaluation} evaluates and analyses the results of the experiment. Finally, Section \ref{Sec: con} concludes the paper.

\section{Related Work}
\label{Sec: Related Work}
\subsection{Backdoor Attacks to Federated Learning}

The concept of backdoor attacks initially emerged in the context of centralized scenarios \cite{BadNets_gu_2019,Blend_chen_2017}. The distributed nature of federated learning introduced new objectives for backdoor attacks, emphasizing effectiveness, stealthiness, and persistence.  Attackers aim to ensure that the backdoored models they upload not only incorporate the malicious functions into the global model after federated aggregation but also remain undetected by backdoor detection methods. Additionally, attackers seek to maintain the presence of the backdoor throughout the iterative process of federated learning, preventing its erosion over time.

Bagdasaryan et al. \cite{backdoorFL_bagdasaryan_2020} proposed pioneering works in deploying backdoor attacks on federated learning. They introduce the Model Replacement Attack, which amplifies the magnitude of backdoor updates proportionally, thereby ensuring the dominance of backdoor parameters in the global model and enhancing the effectiveness of the attack. Furthermore, they propose the Semantic Backdoor Attack, which does not require any modifications to the training samples but instead leverages samples with specific semantic information to activate the backdoor. This approach represents a more stealthy form of backdoor. Inspired by this concept, \cite{wang2020attack} introduces an edge-case backdoor attack that utilizes rare samples (the tail of a dataset) to trigger the backdoor. The Distributed Backdoor Attack (DBA) \cite{DBA_Xie_2020} decomposes a trigger into multiple sub-triggers, with each attacker holding one of these sub-triggers for data poisoning, significantly enhancing the backdoor effectiveness in the global model.

In further developments, more advanced attacks \cite{Neurotoxin_zhang_2022, Towards_Shi_2024, F3BA_Fang_2023} suggest constructing the neuron activation path to unimportant or redundant neurons that are less frequently updated, thereby preventing the backdoor from being promptly erased and enhancing its persistence. Most recently, Li et al. \cite{3DFed_Li_2023} presented 3DFed, which addresses three prominent defense strategies with corresponding attack modules and introduces an indicator mechanism to assess whether backdoor updates are incorporated into model aggregation. This allows for an adaptive adjustment of the attack strategy. We have also observed a category of backdoor attacks based on trigger optimization, such as A3FL \cite{A3FL_Zhang_2024}, F3BA \cite{F3BA_Fang_2023} and CerP  \cite{CerP_Lyu_2023}. These methods aim to obtain a robust trigger to make the attacks more covert and persistent. However, these attacks still operate within the realm of supervised learning, establishing a connection between the backdoor trigger and the target label through loss functions. Despite improvements, the anomaly of backdoored models resulting from this end-to-end training persists. Our method seeks to transcend this limitation.

\subsection{Backdoor Defenses to Federated Learning}

The methods for backdoor attacks in federated learning can be broadly categorized into two main approaches. The first follows the spirit of anomaly detection. The core hypothesis of such methods is that poisoned local models are outliers that deviate significantly from benign models. Consequently, these methods aim to detect outliers by computing the discrepancies between model weights/gradients, model representations, or other metrics, and then exclude abnormal updates to mitigate poisoning attacks. Krum \cite{multikrum_Peva_2017} selects only the local model with the smallest sum of squared Euclidean distances from its $n-m-2$ nearest neighboring models to serve as the global model, where m is an upper bound on the number of malicious participants in FL. Multi-Krum \cite{multikrum_Peva_2017} extends Krum by selecting multiple local models based on the Krum criterion and averaging the selected local models to form the global model. Foolsgold \cite{Foolsgold_Fung_2022} assigns lower aggregation weights to updates with high pairwise cosine similarities, thereby mitigating the impact of backdoor updates. RFLBAT \cite{RFLBAT_Wang_2022} discriminates malicious models according to the difference between poisoned and clean updates in a low-dimensional projection space. Deepsight \cite{DeepSight_Rieger_2022} aims to identify backdoor updates by measuring the fine-grained differences between model updates, generally assuming that the training data of backdoor models exhibits less heterogeneity than that of benign models. BackdoorIndicator \cite{Backdoorindicator_Li_2024} proposed the BackdoorIndicator (referred to as “Indicator” hereafter), which detects potentially poisoned models based on the out-of-distribution (OOD) properties of backdoor samples.

The second approach involves methods for mitigating backdoors, which primarily focus on suppressing or perturbing the magnitude of updates rather than detecting these malicious models. The essence of these methods is to disrupt the clustering of backdoor features in the model’s feature space. Xie et al. \cite{xie2021crfl} propose to apply clipping and smoothing on model parameters to control the global model’s smoothness, which results in a sample-wise robustness certification against backdoors with limited magnitude. FLAME \cite{FLAME_Nguyen_2022} introduces noise to eliminate backdoors based on the concept of differential privacy. RLR \cite{RLR2021} employs robust learning rates to update the global model, applying a negative learning rate to dimensions with significant directional disparities to mitigate the impact of malicious updates. Our method transcends the model gradient or representation anomalies caused by supervised learning and simultaneously conceals the individual clustering of backdoors in the feature space, making our attack more stealthy and persistent.

\section{Problem Formulation}
\label{Sec: Problem}
\subsection{Federated Learning}
Federated Learning enables $N$ clients to train a global model $w$ collaboratively without revealing local datasets. Unlike centralized learning, where local datasets must be collected by a central server before training, FL performs training by uploading the weights of local models $( \{w_k \mid k \in N\} )$  trained on the local dataset  $\mathcal{D}_k = \{(x_{k,i}, y_{k,i})\}_{i=1}^{n_k}$ of size $n_k$ to a parametric server. Specifically, the global training objective is defined as follows:
\begin{equation}
    w = \arg\min_w \sum_{k=1}^{N} \lambda_k L_{k}(w, \mathcal{D}_k),
    \label{EQ1}
\end{equation}
\begin{equation}
    L_k(w,\mathcal{D}_k) =\frac{1}{n_k} \sum_{i=1}^{n_k} f(w, (x_{k,i}, y_{k,i})),
\label{EQ2}
\end{equation}
where $w$ denotes the optimal global model parameters, $L_k(w, \mathcal{D}_k)$ represents the average loss computed over the dataset \(\mathcal{D}_k\) for client \(k\), \((x_{k,i}, y_{k,i})\) denotes the \(i\)-th sample in \(\mathcal{D}_k\), and \(\lambda_k\) indicates the weight of the loss for client \(k\).

At the \(t\)-th federated training round, the server randomly selects a client set \(S_t\) where \(|S_t| = m\) and \(0 < m \leq N\), and broadcasts the current model parameters \(w^t\). The selected clients perform local training in the following three steps:

\begin{itemize}
    \item Global model download. All clients download the global model \( w^t \) from the server.
    \item Local training. Each client updates the global model by training with their datasets: $w_k^t \leftarrow w_k^t - \eta \frac{\partial L(w_k^t,b)}{\partial w_k}, $
    where \( \eta \) and \( b \) refer to learning rate and local batch, respectively.
    \item Aggregation. After the clients upload their local models \( \{w_k^t \mid k \in n\} \), the server updates the global model by aggregating the local models. 
    \begin{equation}
        w^{t+1} \leftarrow w^t + \textbf{AGG}(w_{k}^{t}|k\in S_{t}).
    \label{EQ3}
    \end{equation}
\end{itemize}
    Note that, $\textbf{AGG}(\cdot)$ is the pre-define aggregation method, such as FedAvg \cite{FL_mcmahan_2017}, etc.

\subsection{Threat model}
We build upon the attacks proposed in previous works \cite{backdoorFL_bagdasaryan_2020,DBA_Xie_2020,A3FL_Zhang_2024,PGD_Wang_2020}, where the attackers aim to inject malicious functions, often referred to as hidden neural trojans, into the global model. These neural trojans remain dormant during normal operations but would be activated when specific pre-defined patterns, known as triggers, are present in the input data. Formally, let $ F_{w}$ represent a clean FL model and $\tilde{F}_{w} $ a backdoored model. The objectives of such an attack can be defined as follows:
\begin{equation}
	\begin{split}
\left.\left\{\begin{array}{l}F_{w}(x)=\tilde{F}_{w}(x)\\\tilde{F}_{w}(x\oplus \delta)=y_t\end{array}\right.\right.,
	\end{split}
	\label{EQ4}
\end{equation}
In the context of FL, the attacker has full access to the local training data of compromised clients and can manipulate the entire training process. Existing works typically inject backdoors through end-to-end supervised training. Specifically, given clean local samples $ \mathcal{D}_k $, the attacker seeks to corrupt a fraction $r$ of the data by adding specific triggers $\delta$, thereby constructing a backdoor dataset $ \mathcal{D}_b $, where $\mathcal{D}_k = \mathcal{D}_{c,k} \cup \mathcal{D}_{b,k},  |\mathcal{D}_{b,k}| = r \cdot |\mathcal{D}_k|$. The attacker then trains a backdoor local model $\tilde{F}_{w_{k}} $ by minimizing the following empirical error:
\begin{equation}
	\begin{split}
\mathbb{E}_{(x,y)\thicksim\mathcal{D}_k}[\ell(F_{w_{k}}(x),y)]=\mathbb{E}_{({x},y)\thicksim\mathcal{D}_{c,k}}[\ell(F_{w_{k}}(x),y)]\\+\mathbb{E}_{(x,y_t)\thicksim\mathcal{D}_{b,k}}[\ell(F_{w_{k}}(x),y_t)].
	\end{split}
	\label{EQ5}
\end{equation}
\subsubsection{Adversaries' Capability and Knowledge} The adversary neither has knowledge of nor can manipulate the aggregation rule employed by the central server. Furthermore, the adversary cannot access or modify any information related to benign participants, such as their local datasets or models. The adversary is limited to knowing and controlling only the local training data and processes of the compromised (malicious) participants. To simulate a more realistic scenario and address the limitations of existing studies, which assume the adversary controls multiple clients and engages in continuous backdoor attacks, we narrow our focus to a scenario where the adversary controls only a single client and this client participates in the training for only a limited number of communication rounds.

\section{Methodology}
\label{Sec: Methodology}
\subsection{Attack Intuition and Overview}
\label{Subsec: Attack intuition and challenges}
\subsubsection{Key Intuition}
Consider the backdoor injection attack procedure as outlined in Equation \ref{EQ5}. Typically, the attacker modifies a subset of the training images by embedding triggers and employs an end-to-end training strategy to establish a strong association between the backdoor triggers and the target model. To avoid the feature disentanglement that results from such direct approaches, resulting in the backdoor model exhibiting noticeable deviations from benign models, a more intuitive idea is to directly establish a relationship between the trigger features (embeddings) and the target class feature distribution. In other words, the goal is to trick the model into recognizing the trigger features as normal features of the target class. By integrating the backdoor task into the normal feature learning process, rather than treating it as a separate task, the adversary ensures that the maliciously trained model behaves in a manner indistinguishable from that of benign participants.

\begin{figure*}
	\centering
	\includegraphics[width=1\linewidth]{F2}
	\caption{Workflow of \sysname.}
    {
    }
	\label{F2}
	\vspace{-3mm}
\end{figure*}

\subsubsection{Overview}
As illustrated in Figure \ref{F2}, our approach, denoted as SPA, consists of two key components: \textit{backdoor injection} and \textit{backdoor enhancement}. 
\begin{itemize}
    \item In the backdoor injection module, SPA directly establishes the connection between the trigger feature and the target class feature via feature alignment, so that the feature embeddings of trigger-embedded samples closely resemble those of the target class samples, thereby inducing model misclassification.  Furthermore, we extract clean knowledge from the frozen global model to preserve the utility of the backdoor model.
    \item In the backdoor enhancement module, considering that the fixed trigger feature is difficult to align with the target distribution, we search for adversarial noise from the global model feature space that can inherently steer the predictions of the global model towards the backdoor label as the backdoor trigger, making the backdoor more effective and durable. 
\end{itemize}

\subsection{Backdoor Injection}
\label{Subsec: ID mapping construction}
In the FL setting, each selected participant in round $t$ receives the global model $w_t$ distributed by the server. Following the intuition outlined in Section \ref{Subsec: Attack intuition and challenges}, we do not establish a direct connection between backdoor features and target labels in an end-to-end manner. Instead, we focus on creating a strong association between backdoor features and the feature distributions of the target class. A straightforward and effective approach is to pull the feature embeddings of backdoor samples closer to those of the target class samples, thereby deceiving the model into perceiving the trigger features as belonging to the target class distribution.

Specifically, the attackers filter out the sample subsets $\mathcal{D}_{t}$ whose label is the target class $t$ from their local dataset $\mathcal{D}_{a}$.  $x\oplus \delta$ indicates embedding the trigger $\delta$ to an input sample $x$, referred to as a poisoned sample. To ensure the effectiveness of the backdoor, the attacker manipulates the global model so that it generates strongly similar feature embeddings for any poisoned samples and the target sample, thereby effectively establishing a strong correlation between the trigger features and the features of the target class. Such a correlation would indirectly cause the backdoored model to predict the labels desired by the attacker for trigger-embedding samples. Formally, we denote the feature embedding output of the manipulated global model $w_t$ at $t$-th round as $\tilde{\mathcal{F}}_{w^t}$  and the feature alignment objective is defined as follows:

\begin{equation}
\mathcal{L}_{align} = \frac{1}{|\mathcal{D}_{a}|\cdot|\mathcal{D}_{t}|}\sum_{x\in\mathcal{D}_{a}}\sum_{x_{t}\in\mathcal{D}_{t}}s(\tilde{\mathcal{F}}_{w^t}(x\oplus \delta ),\tilde{\mathcal{F}}_{w^t}(x_{t})).
\label{EQ6}
\end{equation}

Here, \( s(\cdot, \cdot) \) represents a similarity measure between two feature embeddings. Typically, the \( L_2 \)-norm is applied to reduce the distance between backdoor and target feature representations. However, in the FL scenario, the feature space of the global model is constantly evolving, and backdoor embeddings are often high-dimensional and contain some extreme values that are challenging to minimize effectively. Moreover, we lack precise control over whether backdoor features move closer to the target features or the opposite, potentially disrupting the feature embeddings of the clean target class. To address the limitations, we introduce the sliced-Wasserstein distance inspired by \cite{DRUPE_tao_2024,WSBA}, a variant of the Wasserstein distance that is well-suited for measuring high-dimensional data and providing a more robust metric to efficiently handle complex distributions. 

\begin{equation}
\begin{array}{c}
\mathcal{W}_{\text {sliced }}(\mathcal{F}_c, \mathcal{F}_b)=\left(\frac{1}{S} \sum_{s=1}^{S} \int_{0}^{1}\left\|\mathcal{F}_{c}^{s}(z)-\mathcal{F}_{b}^{s}(z)\right\|_{2} d z\right)^{1 / 2}
\end{array}
\label{EQ7}
\end{equation}
where $S$ is the number of one-dimensional directions (denoted by the randomly sampled unit vectors). $F_{c}^s$ and $F_{b}^s$ represent the projections of the clean and poisoned embeddings into one-dimensional data points along the direction of slice $s$, respectively.

Furthermore, we must ensure the utility of the backdoored model, meaning it achieves classification accuracy on clean data comparable to that of the global model, thus enhancing the stealthiness of the backdoor. To achieve this, we employ a distillation-like method to transfer effective knowledge of clean features to the backdoor model, preserving its performance on clean data. Specifically, we freeze the global model obtained in the current training round as a teacher model, denoted as $\mathcal{F}_{w^t}$. During backdoor feature alignment, we constrain the backdoored model to produce embeddings on trigger-free samples similar to those of the teacher model. Formally, we define a utility loss to quantify this objective:
\begin{equation}
\mathcal{L}_{utility} = \frac{1}{|\mathcal{D}_{a}|}\sum_{x\in\mathcal{D}_{a}}s(\tilde{\mathcal{F}}_{w^t}(x ),\mathcal{F}_{w^t}(x)).
\label{EQ8}
\end{equation}
Finally, the backdoor injection process can be formulated as the following optimization problem:
\begin{equation}
\arg\min_{w_{t}} \mathcal{L} = \mathcal{L}_{align} + \lambda \mathcal{L}_{utility},
\label{EQ9}
\end{equation}

\subsection{Backdoor Enhancement}
\label{Subsec: ID mapping enhancement}
Although using feature alignment can make the behavior of backdoored models more consistent with that of benign models, significantly enhancing the stealthiness of malicious participants, this approach also weakens the effectiveness of the backdoor, especially when fixed triggers are employed. In end-to-end learning processes, the fixed trigger is often treated as an independent task. While this characteristic contributes to the abnormality of the backdoored model, it also enables a more straightforward and stable association between the trigger features and the target label. In contrast, it is inherently challenging for a model to classify an out-of-distribution anomaly, such as a fixed trigger, as part of a clean distribution. This difficulty arises because the trigger features often conflict with the normal features of the target class. Furthermore, due to the inclusion of more benign updates during federated learning aggregation, the learned backdoor effects may gradually diminish with continued federated training iterations.

To address these challenges, we propose an adversarial adaptation mechanism for trigger optimization to enhance the effectiveness of backdoor attacks. This method aims to better integrate backdoor triggers into the target feature space, ensuring the backdoor remains robust and persistent even under the ongoing FL process.  Specifically, the attacker has access to the current global model $w_{t}$. Inspired by the concept of adversarial examples, we aim to search for an adversarial noise in the feature space that biases the prediction of the global model toward the target class desired by the attacker. This adversarial noise represents naturally occurring perturbations within the target class of the global model, which is an inherent vulnerability in the global model that can function as a ``natural backdoor". Such a backdoor is robust and difficult to eliminate with the iteration of FL communication, as it aligns closely with the model's intrinsic feature representations. By leveraging feature alignment, the model can be misled into learning these robust backdoor features as part of the normal feature distribution of the target class. The optimization of the backdoor trigger can be formally defined and quantified using the following loss function:
\begin{equation}
\mathcal{L}_{\text{enhance}} = \frac{1}{|\mathcal{D}_a|} \sum_{x \in \mathcal{D}_a} \sum_{i=1}^{C} y_{t,i} \log\left(F_{w^t}(x \oplus \delta)_i\right),
\label{EQ10}
\end{equation}
where $C$ is the number of categories.

In the domain of adversarial machine learning, $L_p$ norms, such as the $L_\infty$ norm, are commonly used to regulate the magnitude of adversarial noise, i.e., $\|\delta\|_\infty\leq\varepsilon$. This approach operates within the input space, specifically at the pixel level, where it limits the size of perturbations to ensure they remain visually imperceptible to human observers. Despite its widespread use, this strategy exhibits inherent limitations, as it may not effectively govern the behavior of the model within the feature space, where higher-level representations are encoded. Furthermore, the fixed $L_p$ constraint lacks flexibility, requiring manual adjustment of the perturbation amplitude threshold (e.g., $\varepsilon$ value) to balance the effectiveness and invisibility of the perturbation. This becomes particularly challenging in the dynamically evolving FL scenarios.

\begin{algorithm}[t]
	\caption{\sysname Algorithm} 
	\label{A1}
	\KwIn{Current global model parameters $w^t$, the data of malicious client $\mathcal{D}_a$.} 
	\KwOut{The backdoored model $w^t$.} 
    \textbf{Malicious Client Execution:} \\
    Freeze teacher model and embeddings: $F_{w^t}$, $\mathcal{F}_{w^t}$ \;

    \tcp{\footnotesize{Backdoor Enhancement Phase}}
    $\delta \leftarrow$ Initialize random perturbation\;
    \For{optimization step $i\in[1,2,\cdots,I]$}
    {
        \For{batch $b_{a}=\{(x,y)\}\in \mathcal{D}_{a},b_{t}=\{(x,y)  \mid y = y_t\}\in \mathcal{D}_{a}$}
        {
            $\mathcal{L}_{enhance} \leftarrow \frac{1}{|b_a|} \sum_{x \in b_a} \sum_{i=1}^{C} y_{t,i} \log(F_{w^t}(x + \delta)_i)$\;
            $\mathcal{L}_{consist} \leftarrow \frac{1}{|b_t|} \sum_{x_t \in b_t} s_{\text{proj}}(\mathcal{F}_{w^t}(x_t), \mathcal{F}_{w^t}(x_t \oplus \delta))$\;
            $\mathcal{L} \leftarrow \mathcal{L}_{enhance} +  \mathcal{L}_{consist}$\;
            $\delta \leftarrow \delta - \eta_{\delta}\nabla_{\delta} \mathcal{L}$\;
        }
    }
    \tcp{\footnotesize{Backdoor Injection Phase}}
    \For{each local epoch $e\in[1,2,\cdots,E_{attak}]$}
    {
        \For{batch $b_{a}=\{(x,y)\}\in \mathcal{D}_{a},b_{t}=\{(x,y)  \mid y = y_t\}\in \mathcal{D}_{a}$}
        {
            $\mathcal{L}_{align} \leftarrow \frac{1}{|b_a|\cdot|b_{t}|}\sum_{x\in b_a}\sum_{x_{t}\in b_t}\mathcal{W}_{\text{sliced}}(\tilde{\mathcal{F}}_{w^t}(x\oplus \delta), \tilde{\mathcal{F}}_{w^t}(x_{t}))$\;
            $\mathcal{L}_{utility} \leftarrow \frac{1}{|b_a|}\sum_{x\in b_a}\mathcal{W}_{\text{sliced}}(\tilde{\mathcal{F}}_{w^t}(x), F_{w^t}(x))$\;
            $\mathcal{L} \leftarrow \mathcal{L}_{align} + \lambda \mathcal{L}_{utility}$\;
            $w^{t} \leftarrow w^{t}-\eta\nabla \mathcal{L}$\;
        }
    }
    \Return{$w^{t}$}
 \vspace{-1mm}
\end{algorithm}

To address these shortcomings, we propose an innovative approach that constrains the magnitude of adversarial perturbations through feature consistency. Specifically, we enforce a condition whereby samples augmented with triggers retain a high degree of similarity to their original counterparts within the feature space. Such strategies enable attackers to dynamically adjust the strength of the backdoor by leveraging the most recent global model, thereby enhancing the stealthiness of the attack while maintaining its efficacy. To quantify this similarity, we adopt the projection distance, a metric that emphasizes the directional alignment of features while remaining insensitive to variations in their magnitudes. The projection distance is formally defined as:

\begin{equation}
s_{\text{proj}}(a, b) = \left\| a - \frac{a \cdot b}{\|b\|^2} b \right\|_2 .
\label{EQ11}
\end{equation}

This measure ensures that the features of perturbed samples are distributed along the principal direction associated with a given class, rather than being forcibly aligned with a specific individual sample. As a result, the constraint imposed on the trigger can be articulated as:
\begin{equation}
\mathcal{L}_{consist}=\frac{1}{|\mathcal{D}_t|} \sum_{x_t \in \mathcal{D}_t} s_{\text{proj}}\left(\mathcal{F}_{w^t}(x_t), \mathcal{F}_{w^t}(x_t \oplus \delta)\right).
\label{EQ12}
\end{equation}

It is imperative to note that we restrict feature alignment to only the target class samples within each batch, rather than applying it universally to all samples. This selective alignment serves a dual purpose: it effectively limits the magnitude of the perturbations, ensuring their subtlety, while simultaneously guiding the trigger features to converge toward the feature distribution characteristic of the target class. By integrating these principles, our approach not only enhances the stealthiness of adversarial perturbations but also ensures their robustness and adaptability in practical attack scenarios.
Finally, the backdoor enhancement process can be formulated as the following optimization problem:
\begin{equation}
\arg\min_{\delta} \mathcal{L} = \mathcal{L}_{enhance} + \mathcal{L}_{consist}.
\label{EQ13}
\end{equation}

Accordingly, the overall \sysname algorithm for an attacker can be summarized in Algorithm \ref{A1}.

\section{Experimental Evaluation}
\label{Sec: Experimental Evaluation}
In this section, we rigorously evaluate the efficacy of \sysname against the SOTA defenses in FL through comprehensive experimental. Our experimental analysis is conducted on real-world datasets within a simulated FL environment. We begin by benchmarking the performance of \sysname against leading backdoor attack techniques under various defense strategies, with particular focus on its persistence and stealth. Next, we measure the effectiveness of \sysname under diverse attack scenarios and FL configurations. Subsequently, we verify the performance of different attack methods on multi-label attacks. Furthermore, we conduct ablation studies to investigate the impact of key parameters on the performance of \sysname, thereby providing deeper insight into the factors contributing to its success.  The empirical results corroborate that \sysname exhibits remarkable efficiency, stealthiness, and resilience, capable of bypassing current defense strategies with ease. To facilitate further research and the development of more sophisticated defenses against such attacks, our implementation is publicly accessible for the reproduction of experimental outcomes.

\textbf{Datasets and Models:} We evaluate \sysname  on three widely-used benchmark datasets: CIFAR-10, CIFAR-100 \cite{Cifar_Krizhevsky_2009}, and GTSRB \cite{GTSRB_houben_2013}. The CIFAR-10 dataset consists of 60,000 images shaped $32\times32$ distributed across 10 classes, where 5,0000 samples are for training and 1,0000 for testing. CIFAR-100 has the same number of images but 100 classes containing 600 images each, with 500 training images and 100 testing images. The German Traffic Sign Recognition Benchmark (GTSRB) contains  39,209 training images and 12,630 test images that are uniformly distributed across 43 classes, where each image has a size of $32\times32$ pixels.

ResNet18 \cite{Resnet_He_2016} is employed as the default model architecture, which is a classic and effective model for image classification. Besides, we also evaluate the performance of \sysname under different model architectures in  Appendix \ref{A-Model_type}, including ResNet34 \cite{Resnet_He_2016}, VGG11, VGG19 \cite{VGG_Simonyan_2015}, and MobileNet-V2 \cite{Mobilenetv2_sandler_2018}.

\textbf{Federated Learning Setup:}
Our experiment implements all the tasks in the FL system running image classification tasks using FedAVG on a single machine using an NVIDIA GeForce RTX 3090 GPU with 24GB of memory. At each communication round, the server randomly selects ten clients to contribute to the global model while the total number of clients is 100. Following previous research \cite{zhu2021data,BadCleaner_JLZ_2024,FLPurifier_JLZ_2024}, we randomly split the dataset over clients in a non-IID manner, using Dirichlet sampling \cite{Dirichlet_hsu_2019} with the parameter $\alpha$ set to 1 by default.  We also evaluate the impact of data heterogeneity by adjusting the value of $\alpha$. During the training process, each selected client trains the local model for two local epochs using the SGD optimizer with a learning rate of 0.01. The batch size is 64 for CIFAR-10 and CIFAR-100, and 32 for GTSRB. The FL training process continues for 2,100 communication rounds.

\textbf{Attack and Defense Setups:} To compare the effectiveness of \sysname with SOTA methods, we conduct experiments against A3FL \cite{A3FL_Zhang_2024}, Chameleon \cite{Chameleon_dai_2023}, Neurotoxin \cite{Neurotoxin_zhang_2022}, PGD \cite{PGD_Wang_2020}, and Vanilla \cite{backdoorFL_bagdasaryan_2020}. Among these methods, A3FL, Chameleon, and Neurotoxin are designed to implant more persistent backdoors. Specifically, A3FL optimizes the trigger pattern to survive scenarios where the global model is explicitly trained to unlearn the trigger, making it less susceptible to removal by global training dynamics. Neurotoxin attempts to inject backdoors by targeting parameters that benign clients rarely update, while Chameleon enhances backdoor persistence by exploiting sample relationships and employing supervised contrastive learning to train the backdoor model. To achieve higher stealth, the Vanilla backdoor attack first constructs malicious training datasets by mixing backdoor samples with benign samples, then trains the backdoor model on this constructed dataset by optimizing cross-entropy loss. In contrast, the PGD backdoor attack employs Projected Gradient Descent (PGD) \cite{PGD_Wang_2020} to train the backdoor model, periodically projecting model parameters onto a sphere centered around the model from the previous iteration, effectively evading norm-clipping defenses designed to mitigate the effects of abnormally large updates.

For trigger-optimization-based backdoor attacks, namely \sysname and A3FL, we configure each attacker to optimize their trigger pattern using PGD \cite{PGD_Wang_2020} with a step size of 0.001. For other attacks, we adhere to their original experimental configurations. To simulate more realistic scenarios, the attack window begins at the 2000th round and continues for 100 rounds, forcing attackers to adapt to random client selection and preventing continuous attack execution. Additionally, we explore the impact of various attack scenarios.

We also evaluate those attacks against six SOTA defenses: Multikrum \cite{multikrum_Peva_2017}, Deepsight \cite{DeepSight_Rieger_2022}, Foolsgold \cite{Foolsgold_Fung_2022}, Rflbat \cite{RFLBAT_Wang_2022}, Flame \cite{FLAME_Nguyen_2022} and Indicator \cite{Backdoorindicator_Li_2024}. For those attacks and defense methods, we follow the original implementations. Note that we selected the random noise dataset as the source of the indicator dataset in Indicator.

\textbf{Evaluation metrics:} Following previous works, we evaluate the performance of \sysname with two metrics: \ding{182} attack success rate (ASR), which is the ratio of backdoored inputs misclassified by the backdoor model as the target labels:
\begin{equation}
\begin{split}
    \text{ASR}=\frac{\#\text{successful attacks}}{\#\text{total trials}},
\end{split}
\label{EASR}
\end{equation}
and \ding{183} the accuracy of the main classification task on normal samples (ACC). Note that we report the mean ASR of all the attackers on the global model at the end of FL. A stealthy and persistent attack in FL is characterized by its ability to maintain the model's original ACC while ensuring that the ASR either remains stable or experiences only minimal degradation throughout the FL iterations. These enable the attack to remain undetected by defense mechanisms while maintaining its effectiveness over extended training periods.

\begin{table*}
\caption{Performance comparison between \sysname and six SOTA defense methods}
\label{Tab: COMP}
\centering
\resizebox{0.95\textwidth}{!}{%
\begin{tabular}{cccccccccccccc}
\toprule
\multirow{2}{*}{\textbf{Dataset}}   & \multirow{2}{*}{\textbf{Attack}} & \multicolumn{2}{c}{\textbf{Vanilla}} & \multicolumn{2}{c}{\textbf{PGD}} & \multicolumn{2}{c}{\textbf{Neurotoxin}} & \multicolumn{2}{c}{\textbf{Chameleon}} & \multicolumn{2}{c}{\textbf{A3FL}} & \multicolumn{2}{c}{\textbf{\sysname}} \\ \cmidrule{3-14} 
                                    &                                  & \textbf{ACC}       & \textbf{ASR}    & \textbf{ACC}    & \textbf{ASR}   & \textbf{ACC}        & \textbf{ASR}      & \textbf{ACC}        & \textbf{ASR}     & \textbf{ACC}    & \textbf{ASR}    & \textbf{ACC}    & \textbf{ASR}    \\ \midrule
\multirow{7}{*}{\textbf{CIFAR-10}}  & \textbf{Nodefense}               & 92.21              & 67.03           & 92.05           & 67.32          & 92.25*              & 63.79             & 91.73               & 48.78            & 92.04           & 98.86*          & \textbf{92.55}  & \textbf{99.91}  \\
                                    & \textbf{Deepsight}               & 92.08              & 65.41           & 92.06           & 60.58          & 92.14               & 54.24             & 92.19               & 36.40             & 92.41*          & 96.73*          & \textbf{92.66}  & \textbf{99.88}  \\
                                    & \textbf{Foolsgold}               & 90.34              & 61.49           & 90.89*          & 62.67          & 90.20                & 76.88             & 90.23               & 42.10             & 90.28           & 98.86*          & \textbf{91.32}  & \textbf{99.14}  \\
                                    & \textbf{Indicator}               & 89.32              & 3.51            & 84.58           & 11.27          & 87.14               & 12.67             & 83.54               & 2.22             & 90.52*          & 53.90*           & \textbf{91.14}  & \textbf{98.31}  \\
                                    & \textbf{Multikrum}               & \textbf{92.45}     & 0.40             & 92.16           & 0.71           & 92.10                & 0.18              & 92.31               & 0.34             & 91.54           & 96.03*          & 92.37*          & \textbf{99.96}  \\
                                    & \textbf{Rflbat}                  & 88.29              & 83.41           & 87.63           & 80.17          & 87.25               & 30.92             & 87.43               & 25.12            & \textbf{89.76}  & 98.59*          & 89.14*          & \textbf{99.29}  \\
                                    & \textbf{Flame}                   & 92.34              & 0.26            & 92.35           & 0.20            & 92.44               & 0.21              & 92.52*              & 0.28             & 92.41           & 98.60*           & \textbf{92.52}  & \textbf{99.92}  \\ \midrule
\multirow{7}{*}{\textbf{CIFAR-100}} & \textbf{Nodefense}               & 71.25              & 71.31           & 70.74           & 77.65          & 71.14               & 74.74             & \textbf{71.91}      & 33.09            & 71.55           & 97.45*          & 71.63*          & \textbf{99.97}  \\
                                    & \textbf{Deepsight}               & 70.98              & 78.22           & 71.03           & 73.44          & 71.17               & 70.62             & 71.69               & 31.23            & 71.72*          & \textbf{99.95}  & \textbf{71.83}  & 99.86*          \\
                                    & \textbf{Foolsgold}               & 70.51              & 72.69           & 70.62           & 85.58          & 70.49               & 75.52             & 71.66*              & 36.01            & 71.02           & 98.62*          & \textbf{71.94}  & \textbf{99.45}  \\
                                    & \textbf{Indicator}               & 69.46              & 0.14            & 60.02           & 0.29           & 66.04               & 0.27              & 61.27               & 0.11             & 69.6*           & 98.28*          & \textbf{70.51}  & \textbf{98.99}  \\
                                    & \textbf{Multikrum}               & 71.66              & 0.11            & 71.94           & 0.15           & 71.65               & 0.13              & 71.76               & 0.20              & \textbf{71.97}  & 94.84*          & 71.95*          & \textbf{99.32}  \\
                                    & \textbf{Rflbat}                  & 63.76              & 51.6            & 63.10            & 58.9           & 64.12               & 64.32             & 62.48               & 3.92             & 64.3*           & 98.64*          & \textbf{68.18}  & \textbf{98.89}  \\
                                    & \textbf{Flame}                   & 72.29*             & 0.15            & 72.18           & 0.13           & 72.05               & 0.10               & 72.12               & 0.12             & 72.02           & 95.20*           & \textbf{72.86}  & \textbf{99.38}  \\ \midrule
\multirow{7}{*}{\textbf{GTSRB}}     & \textbf{Nodefense}               & 96.71              & 98.73           & 96.64           & 98.81          & 96.64               & 98.62             & \textbf{96.81}      & 96.33            & 96.68           & 99.02*          & 96.79*          & \textbf{99.57}  \\
                                    & \textbf{Deepsight}               & 96.41              & 81.89           & 96.78           & 98.80           & \textbf{96.88}      & 98.45             & 96.56               & 97.62            & 96.54           & 98.99*          & 96.83*          & \textbf{99.05}  \\
                                    & \textbf{Foolsgold}               & 96.68              & 98.01           & 96.67           & 98.49          & 95.73               & 97.28             & 96.77*              & 97.05            & 96.66           & 99.37*          & \textbf{96.95}  & \textbf{99.87}  \\
                                    & \textbf{Indicator}               & \textbf{95.14}     & 86.21           & 90.19           & 95.54          & 90.53               & 88.33             & 92.04               & 5.96             & 89.38           & 77.15           & 94.59*          & \textbf{100.00}    \\
                                    & \textbf{Multikrum}               & 96.71*             & 0.01            & 96.67           & 0.00              & 96.51               & 0.00                 & 96.67               & 0.02             & 96.60            & 98.18*          & \textbf{96.97}  & \textbf{98.72}  \\
                                    & \textbf{Rflbat}                  & 94.40               & 97.95           & 95.92*          & 97.71          & 94.52               & 99.36             & 94.29               & 96.88            & 95.10            & 100.00*            & \textbf{96.08}  & \textbf{100.00}    \\
                                    & \textbf{Flame}                   & 96.43              & 0.01            & 96.52           & 0.00              & 96.58*              & 0.01              & 95.92               & 5.95             & 96.47           & 98.54*          & \textbf{96.95}  & \textbf{99.45}  \\ \bottomrule
\end{tabular}
}
\vspace{-3 mm}
\end{table*}

\subsection{Attack Performance}
\label{Subsection: Attack performance}

\textbf{Comparison with SOTA Methods.} Table \ref{Tab: COMP} presents a comprehensive performance comparison between \sysname and five SOTA backdoor attacks against FL under various defense mechanisms. To simulate more challenging and realistic scenarios, we configured the experiments with a single attacker participating intermittently through random selection, with an attack window limited to rounds 2000-2100, meaning the attacker was only selected approximately 10 times throughout the training process. We evaluated performance against six widely adopted backdoor defense methods as well as a baseline scenario without any defense deployment. The best results are highlighted in \textbf{Bold}, while the second-best results are marked with an asterisk $*$. Overall, our proposed method consistently achieved superior or near-superior performance across all datasets, maintaining the ASR exceeding 98$\%$ while successfully circumventing all backdoor defense mechanisms. Moreover, \sysname exhibited minimal impact on ACC, ranking among the top-performing methods in this critical metric.

In the absence of defense mechanisms, existing attack methods remained effective to varying degrees. However, the challenging scenario of a single, randomly selected attacker significantly weakened the performance of Vanilla, PGD, Neurotoxin, and Chameleon, resulting in ASR values below 80$\%$ on CIFAR-10 and CIFAR-100 datasets. Furthermore,  most of these attacks are detectable by Indicator, Multikrum, and Flame defenses, though it should be noted that Indicator achieved this at the cost of some ACC degradation. 

Notably, A3FL emerges as a competitive approach due to its innovative integration of unlearning mechanisms into adversarial training. This optimization enhances trigger persistence, resulting in more concealed and durable backdoor features. Theoretically, Indicator, which detects backdoors by injecting indicator tasks using OOD data, should effectively counter A3FL, as backdoor features typically manifest as OOD features that disrupt indicator tasks. However, we observed defensive instability against A3FL in practice, attributable to scenarios where A3FL's learned backdoor feature space diverges from the indicator tasks' feature space. In contrast, our method maintains alignment with clean features without perturbing other OOD features.

While existing attacks achieve relative stealth through selective neuron manipulation or trigger optimization, they fundamentally rely on supervised training paradigms. This dependency creates distinguishable separation between backdoor and benign features in the latent space, making them detectable by advanced defenses.The breakthrough of \sysname lies in feature alignment that strategically entangles trigger features with normal features to substantially reduce the statistical divergence that defenses typically exploit. A more in-depth visualization analysis is provided in Appendix \ref{A_visual}.

\begin{figure}[t]
\centering
\includegraphics[width=1\linewidth]{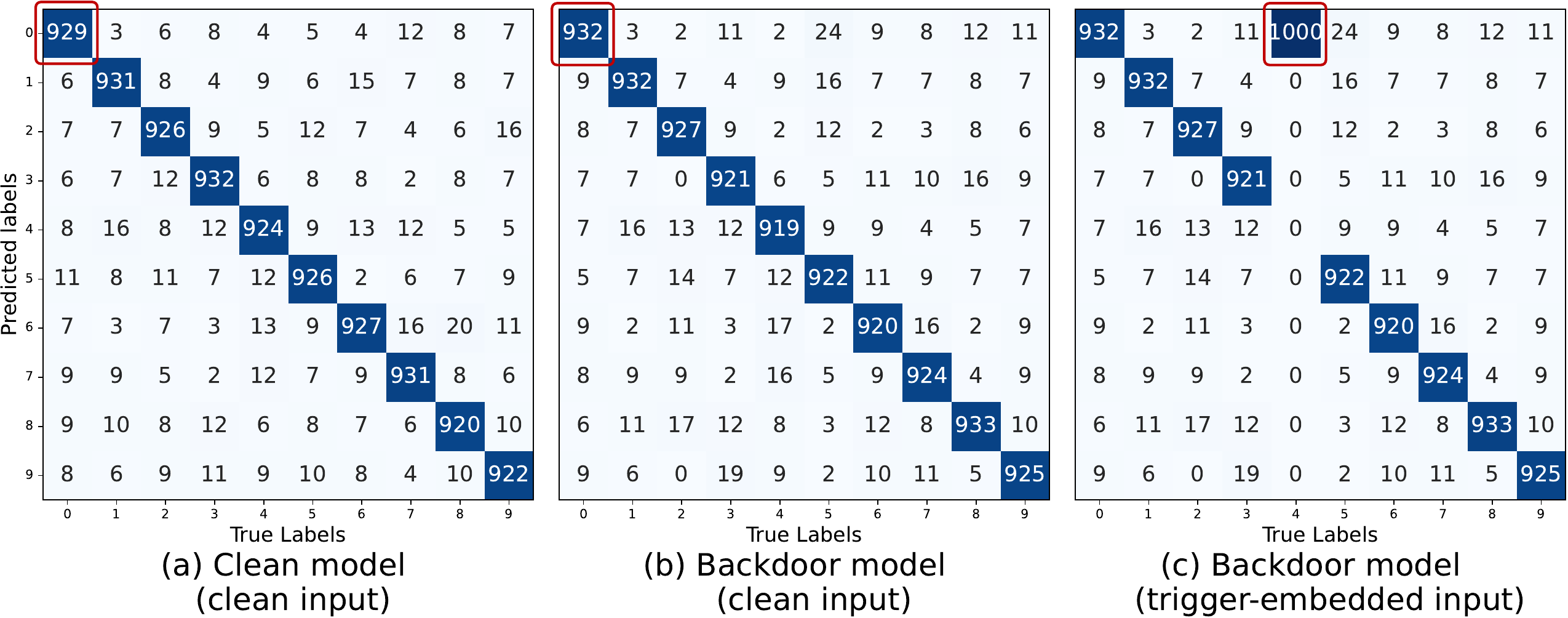}
\caption{Confusion matrix of clean models and backdoor models.}
\label{F3}
\vspace{-3mm}
\end{figure}

\begin{figure}[t]
\centering
\includegraphics[width=0.9\linewidth]{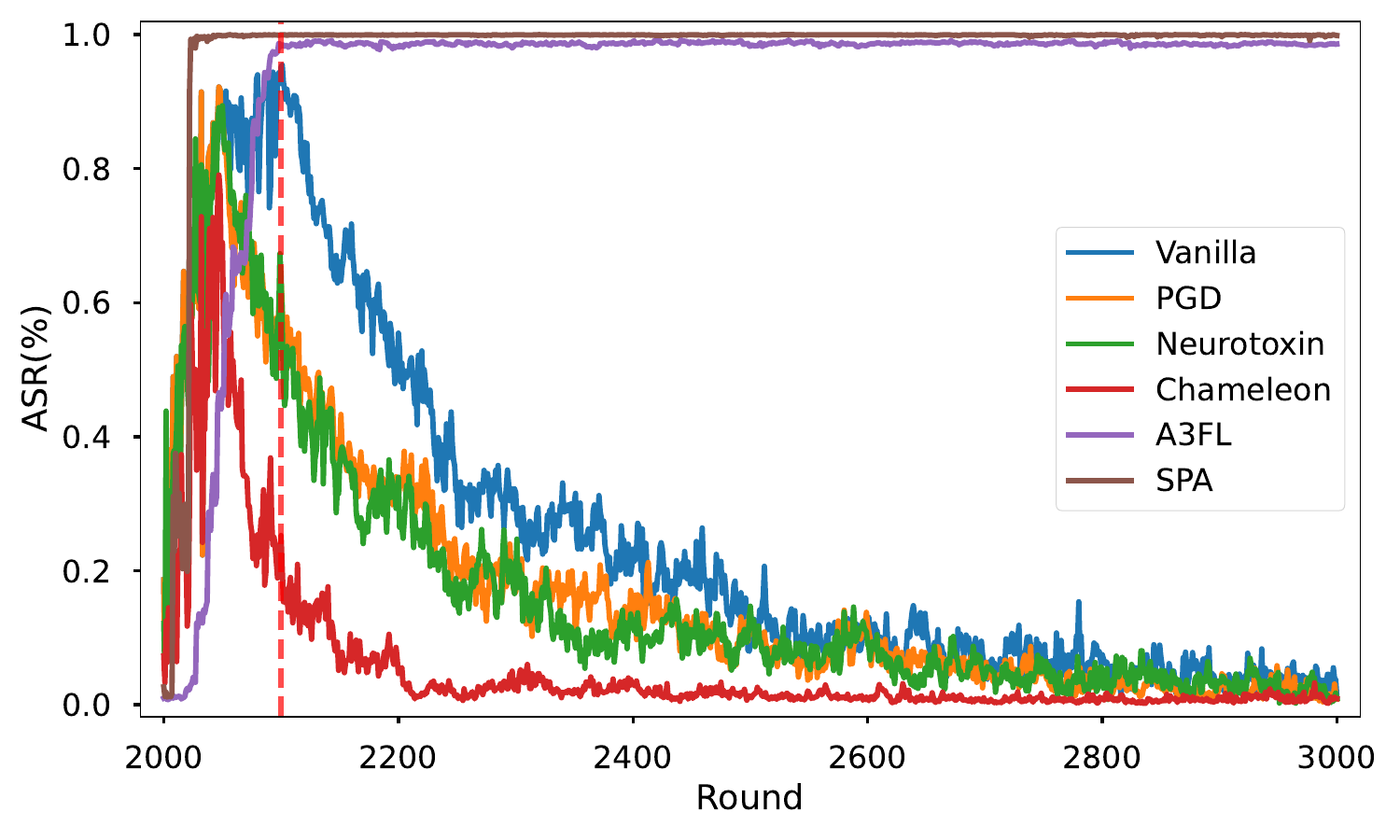}
\caption{Persistence evaluations under no defenses.}
\label{F4}
\vspace{-5mm}
\end{figure}

\begin{figure*}[t]
	\centering
	\subfigure[Persistence evaluations under Deepsight.]{
		\begin{minipage}[t]{0.31\linewidth}
			\centering
			\includegraphics[width=1\linewidth]{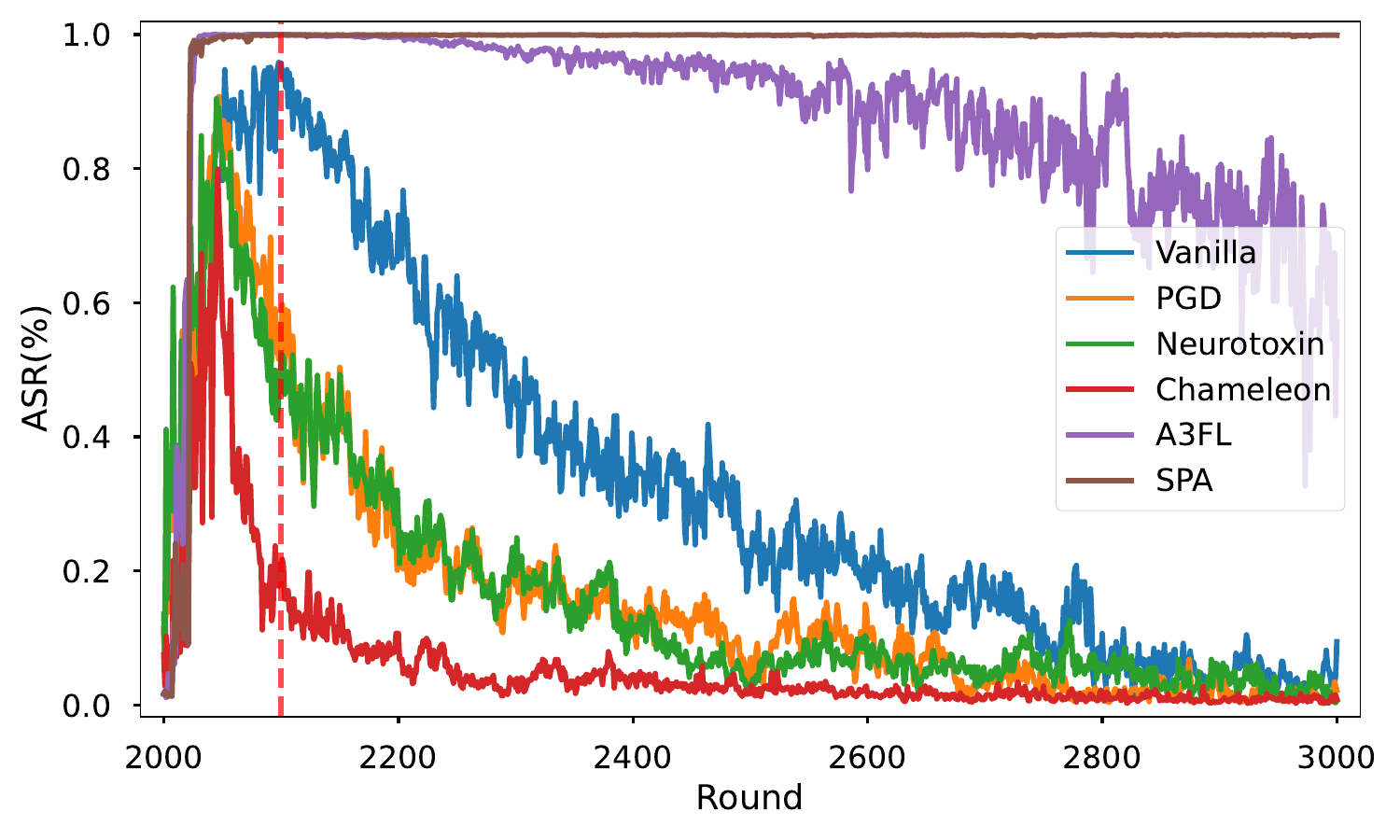}
			\label{F5a}
	\end{minipage}}
 \hspace{2mm}
	\subfigure[Persistence evaluations under Foolsgold.]{
		\begin{minipage}[t]{0.31\linewidth}
			\includegraphics[width=1\linewidth]{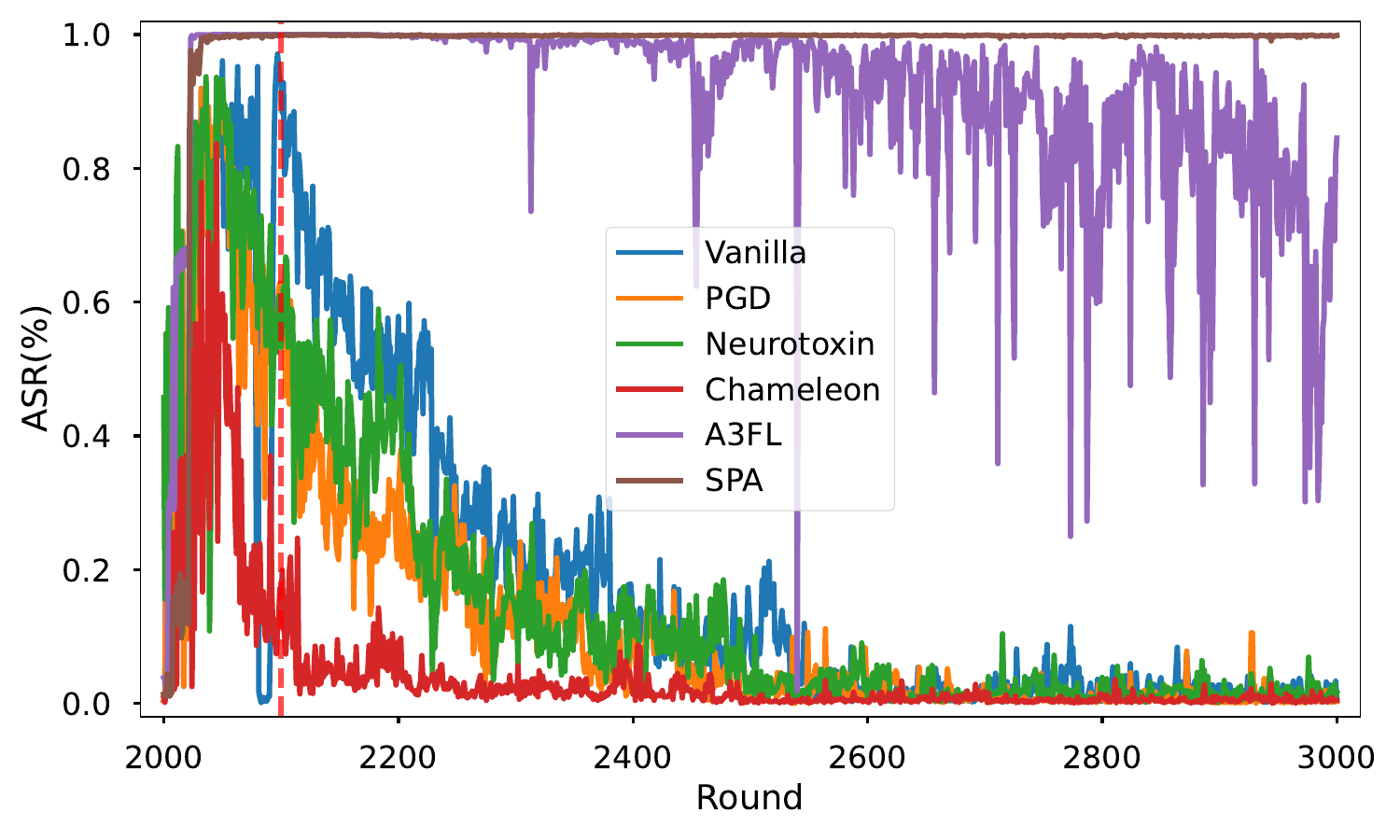}
			\label{F5b}
	\end{minipage}}
  \hspace{2mm}
	\subfigure[Persistence evaluations under Indicator.]{
		\begin{minipage}[t]{0.31\linewidth}
			\centering
			\includegraphics[width=1\linewidth]{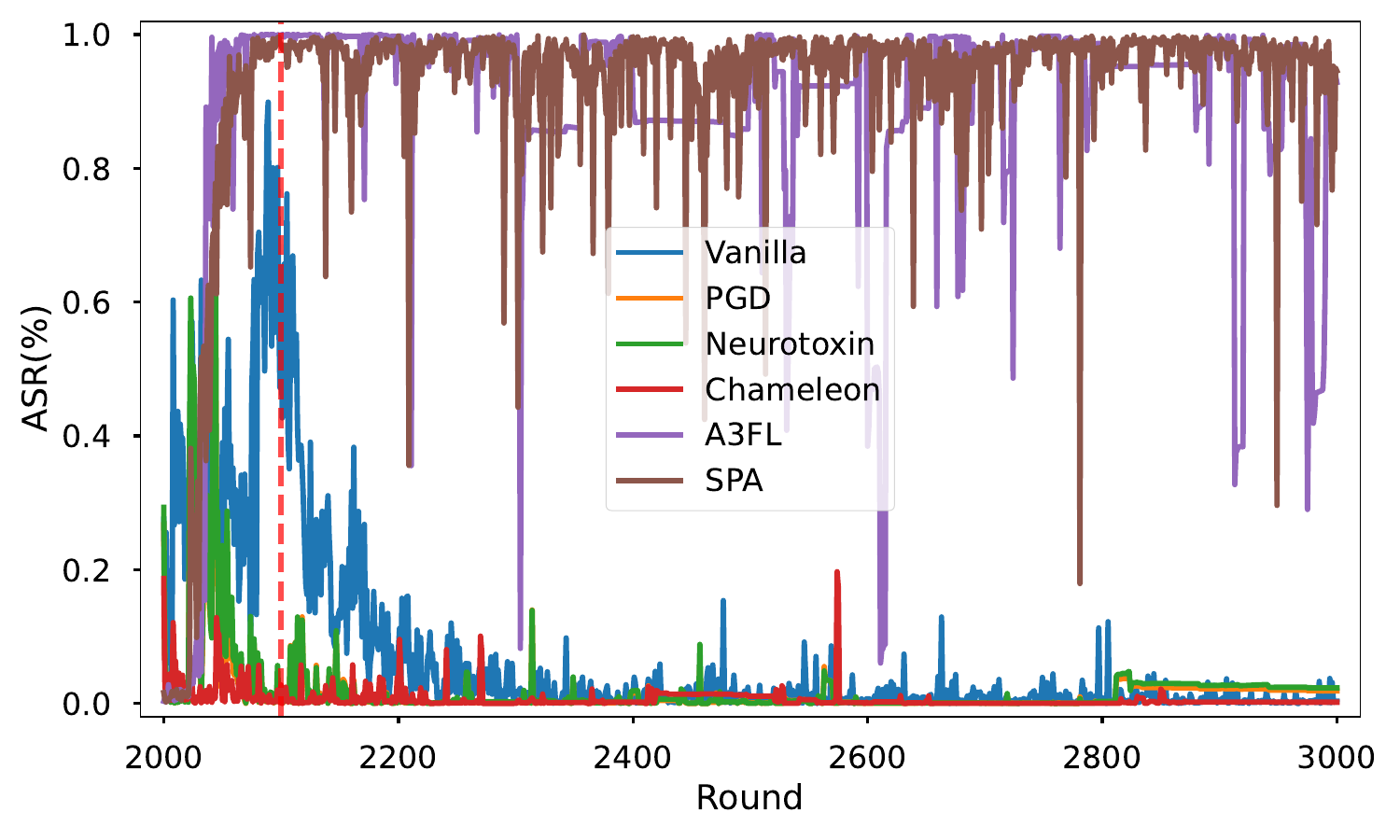}
			\label{F5c}
	\end{minipage}}\\
	\subfigure[Persistence evaluations under Multikrum.]{
		\begin{minipage}[t]{0.31\linewidth}
			\centering
			\includegraphics[width=1\linewidth]{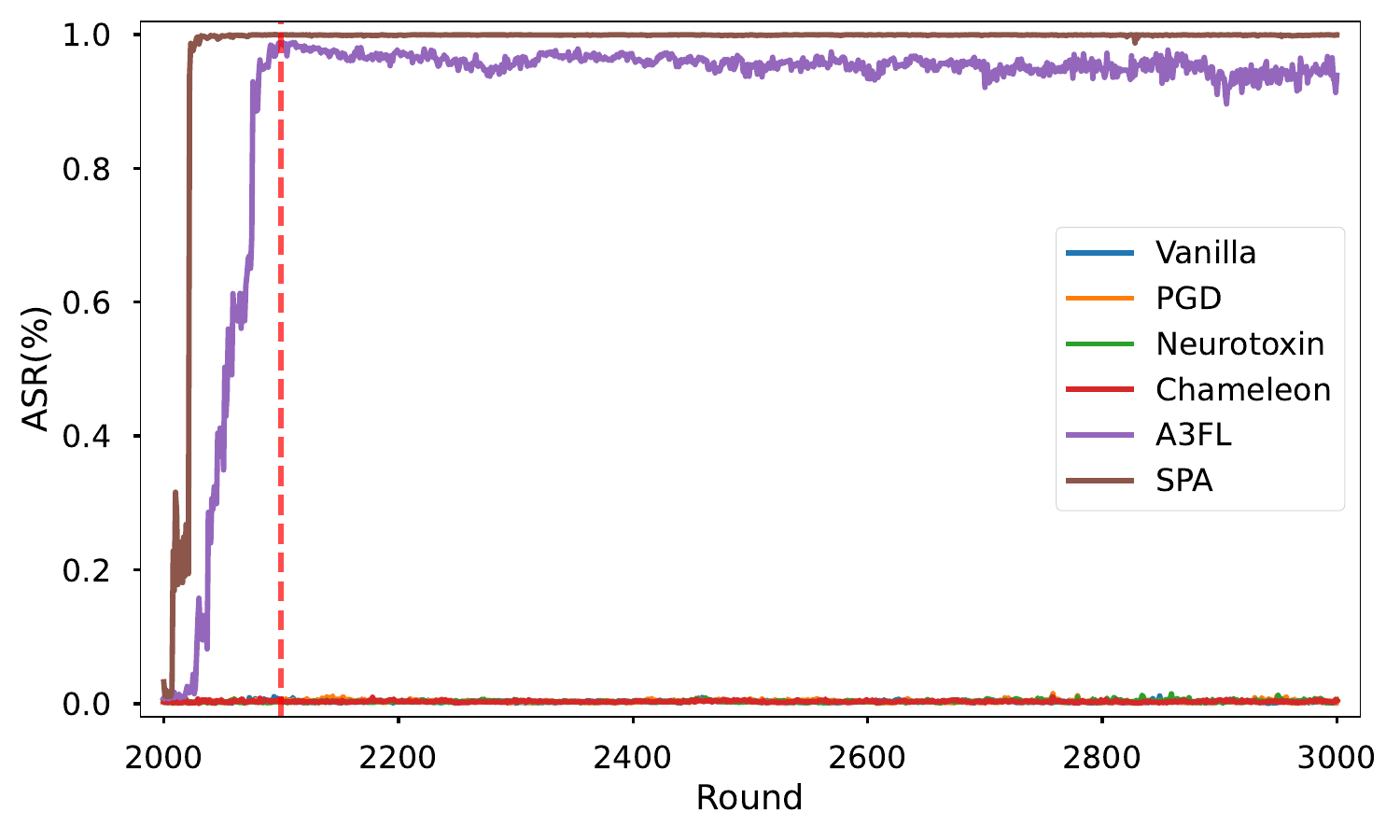}
			\label{F5d}
	\end{minipage}}
  \hspace{2mm}
 \subfigure[Persistence evaluations under Rflbat.]{
		\begin{minipage}[t]{0.31\linewidth}
			\centering
			\includegraphics[width=1\linewidth]{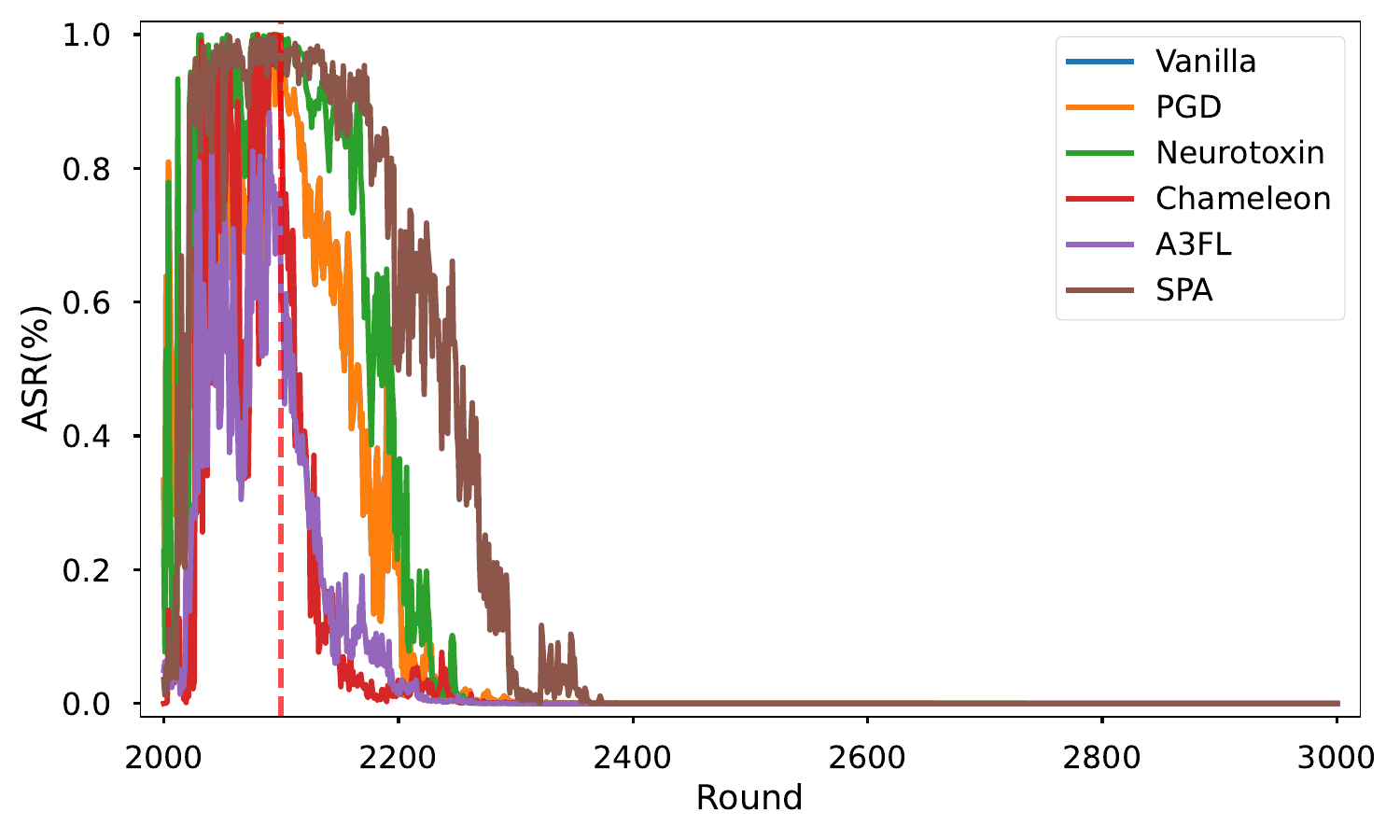}
			\label{F5e}
	\end{minipage}}
  \hspace{2mm}
	\subfigure[Persistence evaluations under Flame.]{
		\begin{minipage}[t]{0.31\linewidth}
			\centering
			\includegraphics[width=1\linewidth]{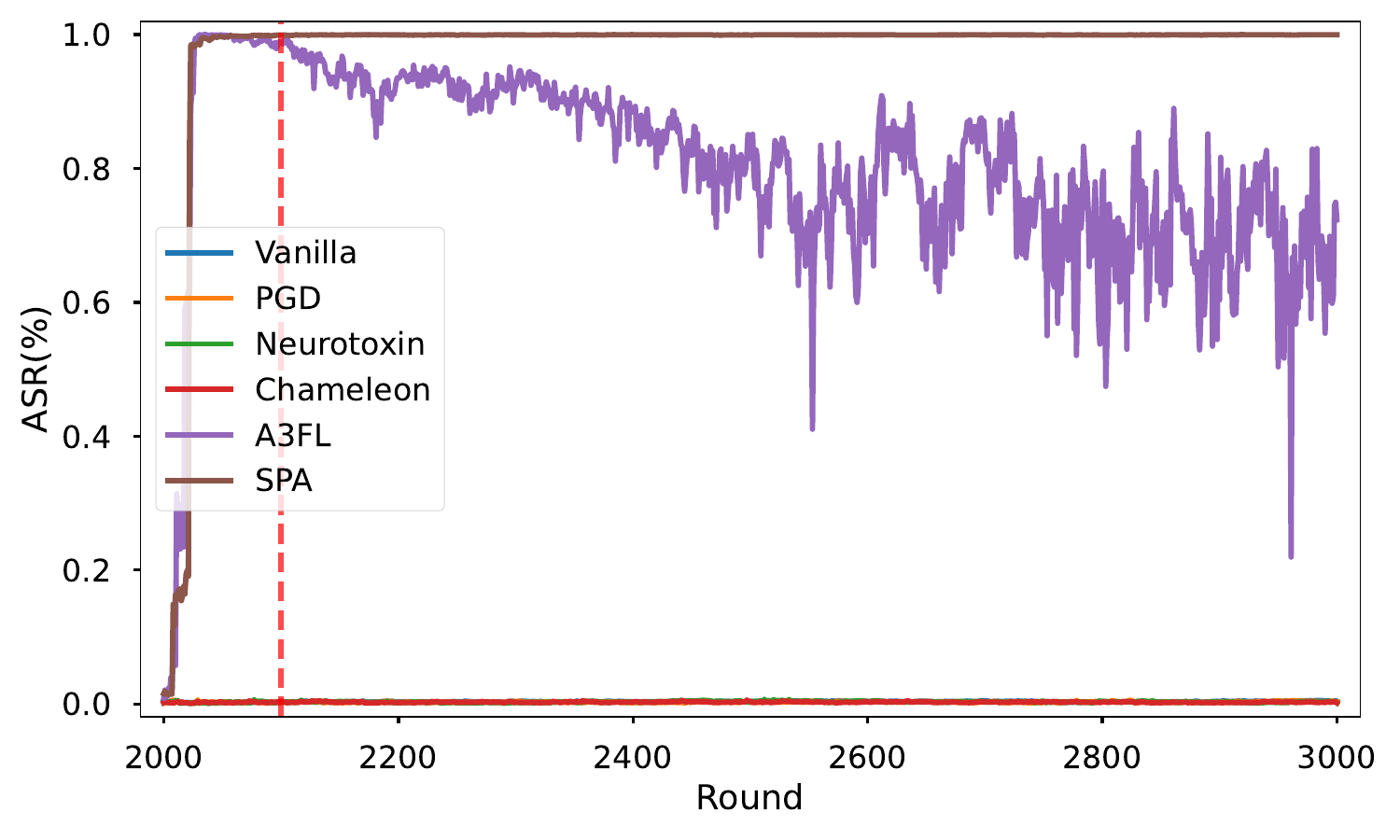}
			\label{F5f}
	\end{minipage}}
	\caption{Comparison of persistence performance under different defense methods.}
	\label{F5}
	 \vspace{-3mm}
\end{figure*}

Furthermore, we investigated the impact of \sysname on the classification accuracy of the target class. Essentially, the core principle of our approach is to mislead the model into perceiving trigger features as legitimate components of the target class features, effectively learning backdoor features as benign features. A potential concern is that achieving high backdoor ASR might cause the model to over-rely on trigger features for target class identification while neglecting other critical features of that class. To systematically evaluate this potential side effect, we examined the confusion matrices before and after the attack to observe any changes in the target class performance. We conducted experiments using the CIFAR-10, which contains 10,000 test samples equally distributed across 10 classes (1,000 samples per class). We designated class 0 as the attacker's target class and applied triggers exclusively to class 5 samples during testing.

As illustrated in Figure \ref{F3}, the comparison between the non-attacked and attacked models reveals minimal difference in the number of correctly classified samples for the target class, where number in Figure \ref{F3}(a) and  Figure \ref{F3}(b) differs by only three. Meanwhile, as shown in  Figure \ref{F3}(c), class 5 samples embedded with triggers were consistently misclassified as class 0.  These results demonstrate that \sysname successfully maintains the model’s original discriminative capability for the target class while achieving strong backdoor efficacy. Crucially, the model does not exhibit over-reliance on trigger features at the expense of learning the natural features of the target class. 


\begin{figure*}[h]
	\centering
	\subfigure[Results on CIFAR-10 dataset.]{
		\begin{minipage}[t]{0.322\linewidth}
			\centering
			\includegraphics[width=1\linewidth]{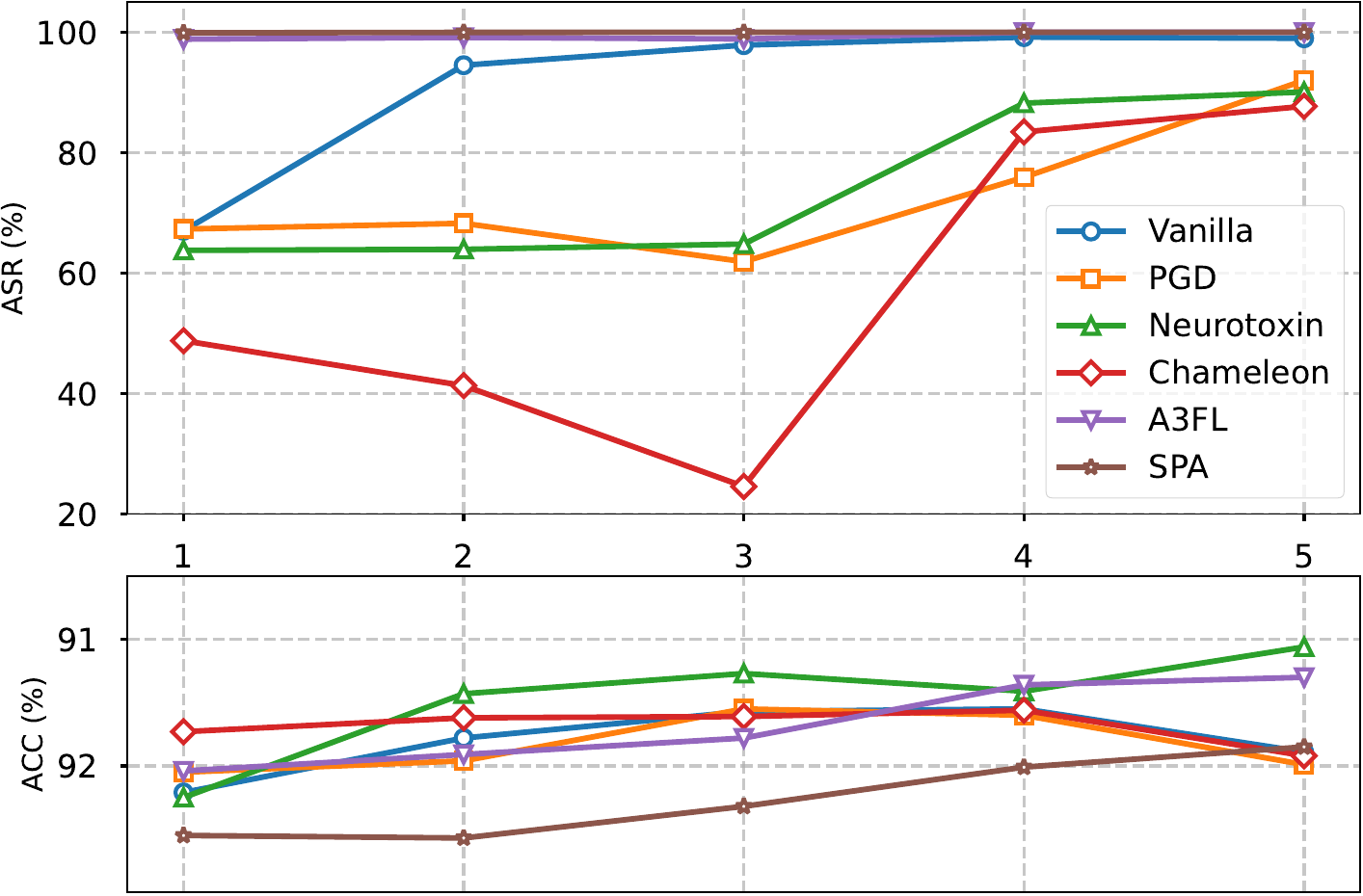}
			\label{F6A}
	\end{minipage}}
	\subfigure[Results on CIFAR-100 dataset.]{
		\begin{minipage}[t]{0.322\linewidth}
			\includegraphics[width=1\linewidth]{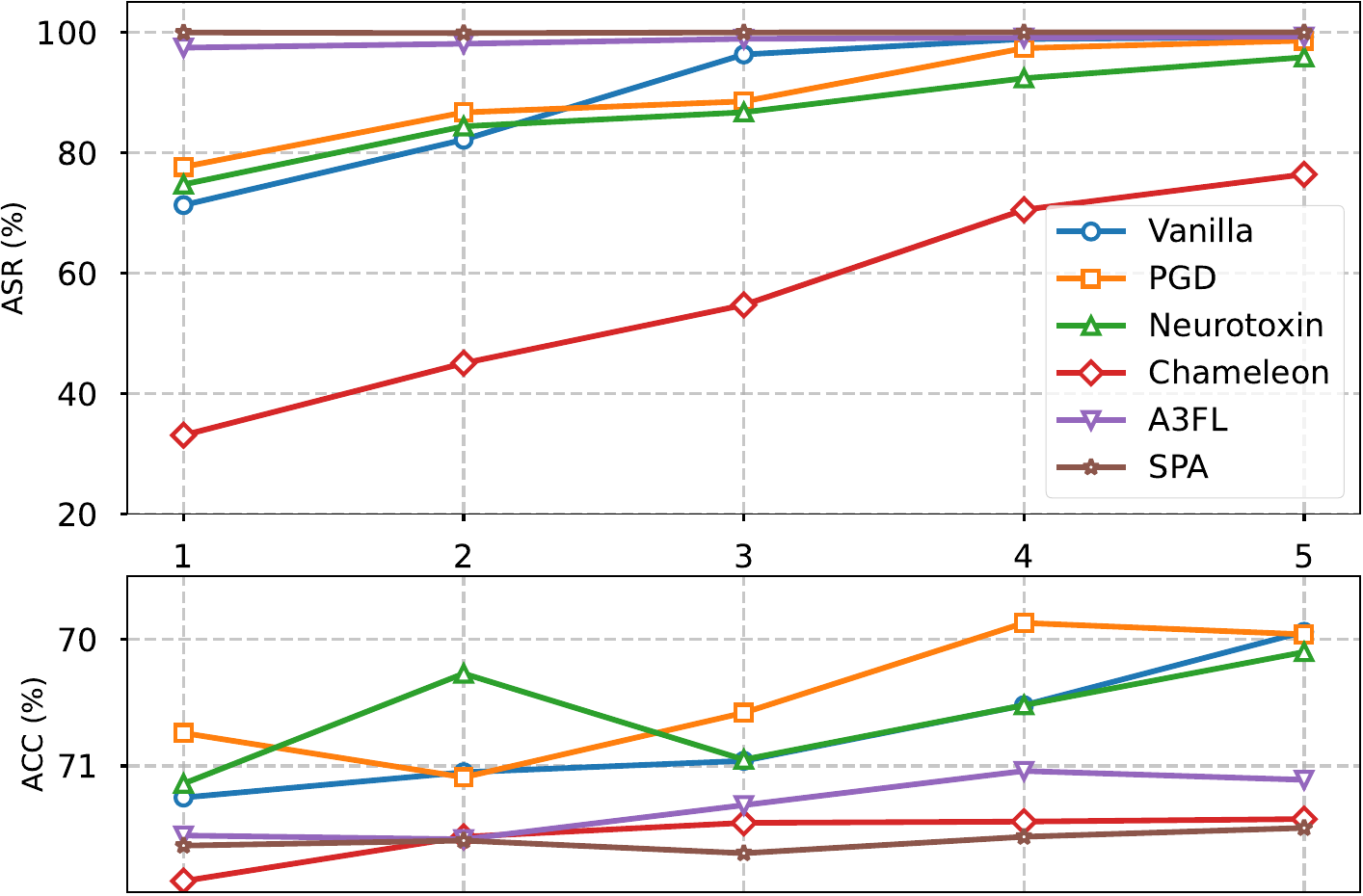}
			\label{F6B}
	\end{minipage}}
	\subfigure[Results on GTSRB dataset.]{
		\begin{minipage}[t]{0.322\linewidth}
			\centering
			\includegraphics[width=1\linewidth]{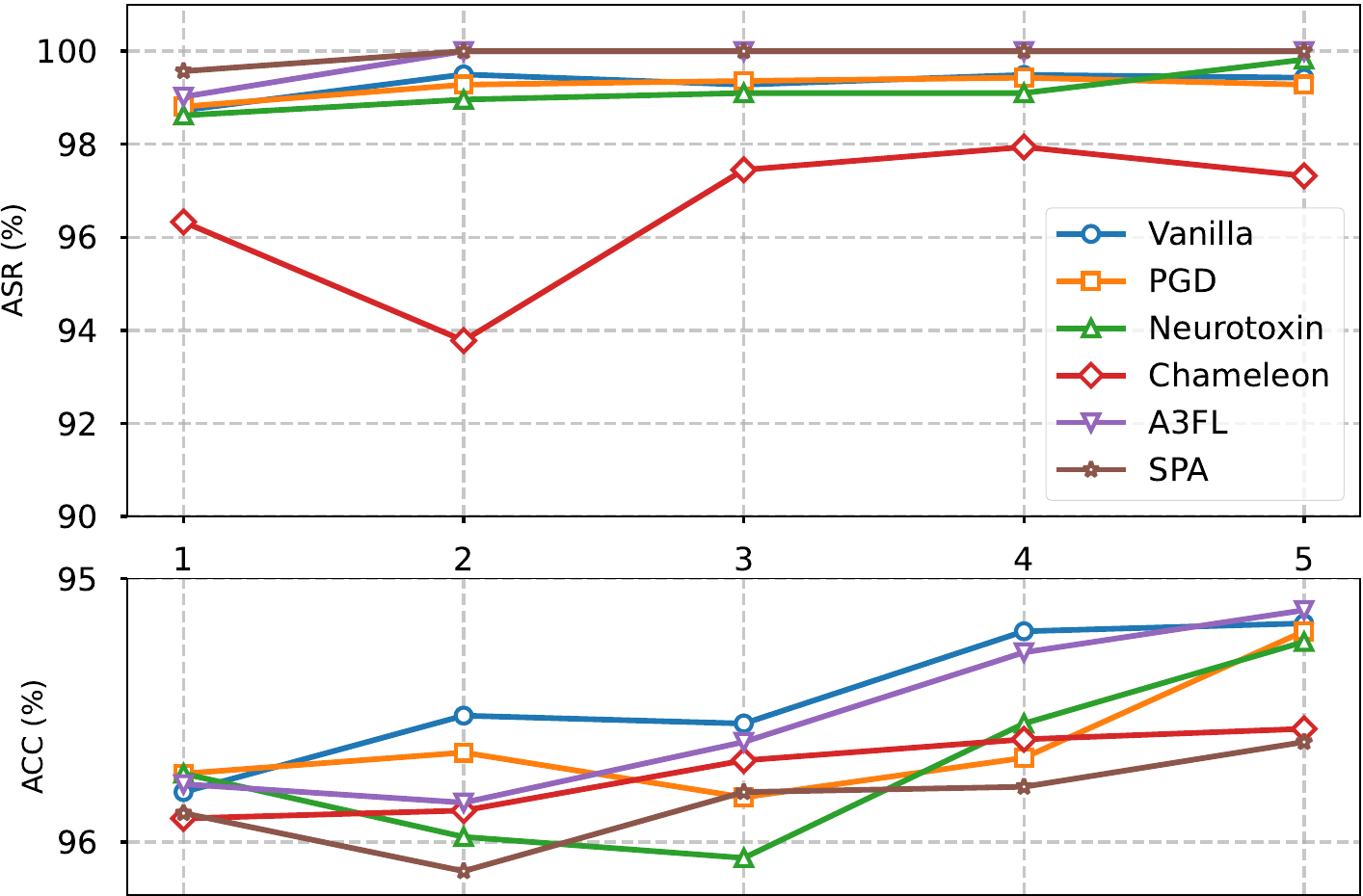}
			\label{F6C}
	\end{minipage}}
	\caption{Comparison of performance under a varying number of malicious clients on three datasets.}
	\label{F6}
	\vspace{-5mm}
\end{figure*}

\textbf{Persistence Analysis.} We further assessed the durability of \sysname by comparing its lifespan with baseline methods. The attack window remained consistent at rounds 2000-2100. To enhance backdoor ASR during this phase, we configured three malicious clients (still selected randomly)  and extended the FL process to 3000 rounds. Figure \ref{F5} illustrates the effect in scenarios without any defense mechanisms. We observed that \sysname demonstrates significantly longer lifespan than baseline approaches, maintaining over 90$\%$ ASR for more than 900 rounds after the attack window. In contrast, baseline attacks, except A3FL, experienced rapid ASR degradation. This phenomenon can be attributed to the inherent limitations of conventional end-to-end backdoor methods: their abnormal gradient patterns during backdoor model updates become diluted during aggregation with benign gradients, leading to progressive attack mitigation. Our approach fundamentally differs by embedding backdoor features as supplementary knowledge within the feature space of target classes, which remains consistent with the core objective of FL that aggregates distributed knowledge from participants. 

A3FL demonstrates comparable persistence through its adversarial training framework combined with unlearning mechanisms. However, its adversarial process incurs substantial computational overhead, representing a significant practical limitation. Moreover, comprehensive experiments across various defense mechanisms illustrated in Figure \ref{F6}  further confirm the superior durability of \sysname, consistently maintaining high attack effectiveness regardless of the deployed defensive strategies. These results collectively demonstrate that our feature-space alignment strategy achieves both stealth and persistence by naturally integrating backdoor features as complementary knowledge rather than conflicting objectives.

\subsection{Applicability Analysis}

\begin{table}
\centering
\caption{Results on different non-IID settings.}
\resizebox{0.46\textwidth}{!}{%
\begin{tabular}{ccccccc}
\toprule
\multicolumn{2}{c}{\textbf{Dataset}}               & \textbf{0.5} & \textbf{1} & \textbf{5} & \textbf{10} & \textbf{1000} \\ \midrule
\multirow{2}{*}{\textbf{CIFAR 10}}  & \textbf{ACC} & 91.56        & 92.55      & 92.62      & 92.59       & 92.88         \\
                                    & \textbf{ASR} & 99.65        & 99.91      & 99.42      & 99.96       & 100.00           \\ \midrule
\multirow{2}{*}{\textbf{CIFAR 100}} & \textbf{ACC} & 71.59        & 71.63      & 71.68      & 71.78       & 71.82         \\
                                    & \textbf{ASR} & 99.24        & 99.97      & 100.00        & 99.76       & 99.85         \\ \midrule
\multirow{2}{*}{\textbf{GTSRB}}     & \textbf{ACC} & 96.28        & 96.79      & 96.81      & 96.91       & 96.86         \\
                                    & \textbf{ASR} & 99.46        & 99.57      & 99.53      & 100.00         & 100.00           \\ \bottomrule
\end{tabular}
}
\vspace{-3 mm}
\label{Tab: different data distribution}
\end{table}

\textbf{Non-IID data distributions.} In FL scenarios, diverse data distributions represent a crucial and realistic setting where participant data is typically distributed in a non-IID manner. Following existing work \cite{Dirichlet_hsu_2019,FLPurifier_JLZ_2024}, we employ the Dirichlet distribution Dir($\alpha$) to model varying degrees of data heterogeneity, where smaller $\alpha$ values correspond to higher data heterogeneity. Specifically, we varied $\alpha$ across values of 0.5, 1, 5, 10, and 1000 across three datasets, with $\alpha=10$ representing the IID scenario. 

As demonstrated in Table \ref{Tab: different data distribution}, \sysname exhibits remarkable robustness against varying data distributions, maintaining consistent performance across all tested conditions. The primary challenge introduced by non-IID lies in the reduced availability of target class samples for potential attackers. Nevertheless, our trigger enhancement mechanism enables effective backdoor feature learning even when target class samples are scarce. This resilience stems from the nature of our optimized trigger features, which inherently represent noise features near the target class decision boundary, thereby facilitating easier feature alignment with the target class using a few target samples.

\textbf{Impact of Adversary Number.} To evaluate the practical applicability of our method in real-world scenarios where attackers typically cannot control multiple participants, we investigate the impact of varying numbers of malicious clients on attack effectiveness. Our experiments systematically assess different attack methods across three benchmark datasets with the number of attackers ranging from 1 to 5, while maintaining all other parameters at their default configurations. 

The experimental results, as illustrated in Figure \ref{F6}, reveal several important insights. Generally, as the number of malicious participants under attacker control increases, the ASR of backdoor attacks on the global model also increases, accompanied by a modest decline in ACC. Notably, \sysname achieved an ASR exceeding 90$\%$ with just a single malicious participant, whereas most baseline attack methods achieved only approximately 70$\%$  ASR under the same conditions. This superior performance demonstrates that our approach is significantly more applicable to real-world scenarios where controlling multiple participants is impractical.

\subsection{Different Attack Settings}

\begin{table*}
\caption{Performance of \sysname compared with baseline attack under multi-label attack scenario.}
\centering
\resizebox{0.9\textwidth}{!}{%
\begin{tabular}{cccccccccccccc}
\toprule
\multicolumn{2}{c}{\multirow{2}{*}{\textbf{Attack}}}    & \multicolumn{2}{c}{\textbf{Vanilla}} & \multicolumn{2}{c}{\textbf{PGD}}  & \multicolumn{2}{c}{\textbf{Neurotoxin}} & \multicolumn{2}{c}{\textbf{Chameleon}} & \multicolumn{2}{c}{\textbf{A3FL}} & \multicolumn{2}{c}{\textbf{SPA}}  \\ \cmidrule{3-14} 
\multicolumn{2}{c}{}                                    & \textbf{ACC}         & \textbf{ASR}  & \textbf{ACC}       & \textbf{ASR} & \textbf{ACC}          & \textbf{ASR}    & \textbf{ACC}          & \textbf{ASR}   & \textbf{ACC}       & \textbf{ASR} & \textbf{ACC}       & \textbf{ASR} \\ \midrule
\multirow{4}{*}{\textbf{Nodefense}} & \textbf{Attack 1} & \multirow{3}{*}{-}   & 0.26          & \multirow{3}{*}{-} & 0.04         & \multirow{3}{*}{-}    & 1.21            & \multirow{3}{*}{-}    & 10.80          & \multirow{3}{*}{-} & 90.44        & \multirow{3}{*}{-} & 99.99        \\
                                    & \textbf{Attack 2} &                      & 7.44          &                    & 92.82        &                       & 66.27           &                       & 17.38          &                    & 99.89        &                    & 99.86        \\
                                    & \textbf{Attack 3} &                      & 87.94         &                    & 0.90         &                       & 91.50           &                       & 90.78          &                    & 96.51        &                    & 99.88        \\
                                    & \textbf{Average}  & 91.76                & 31.88         & 91.52              & 31.25        & 92.03                 & 52.99           & 92.35                 & 39.65          & 91.73              & 95.61        & 92.39              & 99.91        \\ \midrule
\multirow{4}{*}{\textbf{Deepsight}} & \textbf{Attack 1} & \multirow{3}{*}{-}   & 0.52          & \multirow{3}{*}{-} & 0.03         & \multirow{3}{*}{-}    & 2.41            & \multirow{3}{*}{-}    & 86.73          & \multirow{3}{*}{-} & 99.88        & \multirow{3}{*}{-} & 99.85        \\
                                    & \textbf{Attack 2} &                      & 5.80          &                    & 92.66        &                       & 29.40           &                       & 15.34          &                    & 92.14        &                    & 98.76        \\
                                    & \textbf{Attack 3} &                      & 88.81         &                    & 0.56         &                       & 85.91           &                       & 7.91           &                    & 97.66        &                    & 97.52        \\
                                    & \textbf{Average}  & 91.44                & 31.71         & 90.81              & 31.08        & 92.31                 & 39.24           & 92.35                 & 36.66          & 92.17              & 96.56        & 92.46              & 98.71        \\ \midrule
\multirow{4}{*}{\textbf{Foolsgold}} & \textbf{Attack 1} & \multirow{3}{*}{-}   & 1.31          & \multirow{3}{*}{-} & 0.04         & \multirow{3}{*}{-}    & 73.26           & \multirow{3}{*}{-}    & 10.07          & \multirow{3}{*}{-} & 99.60        & \multirow{3}{*}{-} & 98.33        \\
                                    & \textbf{Attack 2} &                      & 5.37          &                    & 94.00        &                       & 95.09           &                       & 77.32          &                    & 98.71        &                    & 98.96        \\
                                    & \textbf{Attack 3} &                      & 90.81         &                    & 0.97         &                       & 74.39           &                       & 48.99          &                    & 87.58        &                    & 96.38        \\
                                    & \textbf{Average}  & 88.79                & 32.50         & 90.71              & 31.67        & 91.04                 & 80.91           & 90.66                 & 45.46          & 89.00              & 95.30        & 91.54              & 97.89        \\ \midrule
\multirow{4}{*}{\textbf{Indicator}} & \textbf{Attack 1} & \multirow{3}{*}{-}   & 0.91          & \multirow{3}{*}{-} & 0.24         & \multirow{3}{*}{-}    & 49.74           & \multirow{3}{*}{-}    & 21.62          & \multirow{3}{*}{-} & 99.96        & \multirow{3}{*}{-} & 97.77        \\
                                    & \textbf{Attack 2} &                      & 0.02          &                    & 1.83         &                       & 0.26            &                       & 2.04           &                    & 11.51        &                    & 99.65        \\
                                    & \textbf{Attack 3} &                      & 85.34         &                    & 57.74        &                       & 16.02           &                       & 0.97           &                    & 99.53        &                    & 97.36        \\
                                    & \textbf{Average}  & 85.08                & 28.76         & 85.83              & 19.94        & 85.13                 & 22.01           & 86.13                 & 8.21           & 85.75              & 70.33        & 91.23              & 98.26        \\ \midrule
\multirow{4}{*}{\textbf{Multikrum}} & \textbf{Attack 1} & \multirow{3}{*}{-}   & 0.16          & \multirow{3}{*}{-} & 0.27         & \multirow{3}{*}{-}    & 0.38            & \multirow{3}{*}{-}    & 0.32           & \multirow{3}{*}{-} & 99.80        & \multirow{3}{*}{-} & 99.65        \\
                                    & \textbf{Attack 2} &                      & 0.03          &                    & 0.01         &                       & 0.06            &                       & 0.07           &                    & 35.39        &                    & 85.82        \\
                                    & \textbf{Attack 3} &                      & 4.59          &                    & 3.26         &                       & 5.24            &                       & 5.32           &                    & 99.27        &                    & 98.72        \\
                                    & \textbf{Average}  & 92.36                & 1.59          & 92.39              & 1.18         & 92.64                 & 1.89            & 92.46                 & 1.90           & 92.18              & 78.15        & 92.58              & 94.73        \\ \midrule
\multirow{4}{*}{\textbf{Rflbat}}    & \textbf{Attack 1} & \multirow{3}{*}{-}   & 0.29          & \multirow{3}{*}{-} & 0.00         & \multirow{3}{*}{-}    & 0.02            & \multirow{3}{*}{-}    & 6.57           & \multirow{3}{*}{-} & 65.13        & \multirow{3}{*}{-} & 98.72        \\
                                    & \textbf{Attack 2} &                      & 2.74          &                    & 99.79        &                       & 0.06            &                       & 11.89          &                    & 99.90        &                    & 96.42        \\
                                    & \textbf{Attack 3} &                      & 88.63         &                    & 0.03         &                       & 99.73           &                       & 92.99          &                    & 98.46        &                    & 98.62        \\
                                    & \textbf{Average}  & 86.83                & 30.55         & 88.28              & 33.27        & 87.22                 & 33.27           & 86.97                 & 37.15          & 85.46              & 87.83        & 88.73              & 97.92        \\ \midrule
\multirow{4}{*}{\textbf{Flame}}     & \textbf{Attack 1} & \multirow{3}{*}{-}   & 0.18          & \multirow{3}{*}{-} & 0.23         & \multirow{3}{*}{-}    & 1.18            & \multirow{3}{*}{-}    & 0.30           & \multirow{3}{*}{-} & 99.54        & \multirow{3}{*}{-} & 98.82        \\
                                    & \textbf{Attack 2} &                      & 0.03          &                    & 0.02         &                       & 26.50           &                       & 0.07           &                    & 39.31        &                    & 99.06        \\
                                    & \textbf{Attack 3} &                      & 5.29          &                    & 9.92         &                       & 95.22           &                       & 3.78           &                    & 99.36        &                    & 99.00        \\
                                    & \textbf{Average}  & 92.56                & 1.83          & 92.51              & 3.39         & 92.02                 & 40.97           & 92.35                 & 1.38           & 92.42              & 79.40        & 92.54              & 98.96        \\ \bottomrule
\end{tabular}

}
\label{Tab: multi}
\vspace{-3 mm}
\end{table*}

\textbf{Multi-label Attack Scenario.} In traditional FL backdoor attacks, a single attacker typically controls multiple clients with a consistent target label, essentially launching a collusion attack. However, real-world scenarios often involve multiple attackers controlling different clients and targeting different labels, resulting in multi-label attacks. In this section, we evaluate the performance of various attack methods in multi-label attack scenarios. Specifically, we configured three attackers with distinct target labels: 0, 1, and 2. We report the average performance across all three attackers.

\begin{figure}[t]
\centering
\includegraphics[width=1\linewidth]{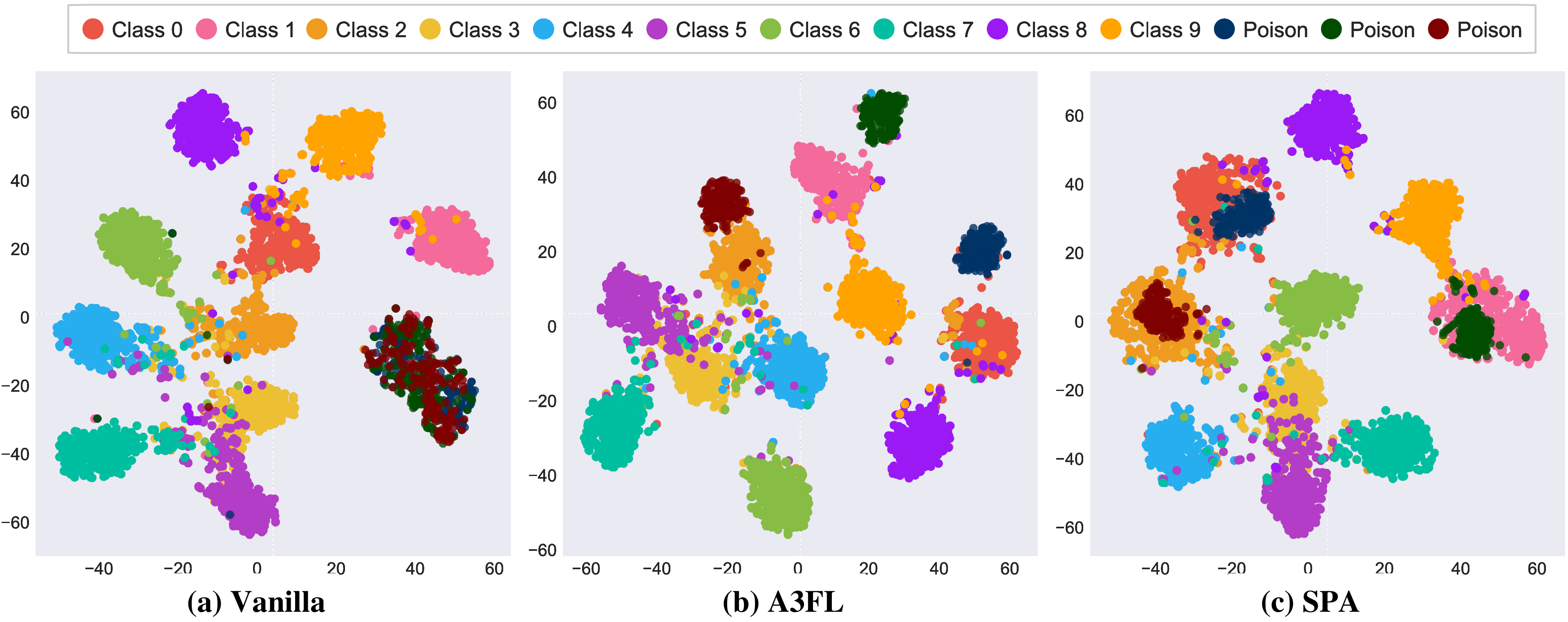}
\caption{T-SNE visualization of features under multi-label scenarios.}
\label{F7}
\vspace{-3mm}
\end{figure}

As shown in Table \ref{Tab: multi}, the experimental results reveal significant performance disparities among different methods. Vanilla, PGD, Neurotoxin, and Chameleon occasionally enabled one attacker to achieve an ASR exceeding 90$\%$, yet their average ASR remained relatively low. This pattern indicates that among the three attackers, only one could successfully compromise the model, while the others failed. In contrast, both A3FL and \sysname enabled all three attackers to achieve high ASR simultaneously. To further investigate this phenomenon, we visualized the T-SNE projections of feature representations for Vanilla, A3FL, and our method, as illustrated in Figure \ref{F7}. The visualization provides critical insights into the underlying mechanisms. For Vanilla attacks, despite having different targets, all attackers employ end-to-end training approaches, causing different out-of-distribution tasks to interfere with each other within the same feature space. This interference explains why Indicator-based defenses can successfully detect backdoors by leveraging OOD data characteristics.

A3FL demonstrates improved performance through trigger optimization, which effectively separates different OOD tasks, enabling successful multi-label backdoor attacks. However, its multiple out-of-distribution backdoor features still result in at least one attacker being detectable, as certain backdoor tasks inevitably interfere with the OOD indicator tasks planted by defense mechanisms. \sysname achieves superior performance by aligning different backdoor features with their respective target class features. This alignment strategy effectively distributes backdoor features across their corresponding target class distributions, simultaneously avoiding inter-attack conflicts and evading detection. By embedding each backdoor within the natural feature distribution of its target class, \sysname creates multiple covert channels that operate independently without mutual interference.

\begin{figure}[t]
\centering
\includegraphics[width=1\linewidth]{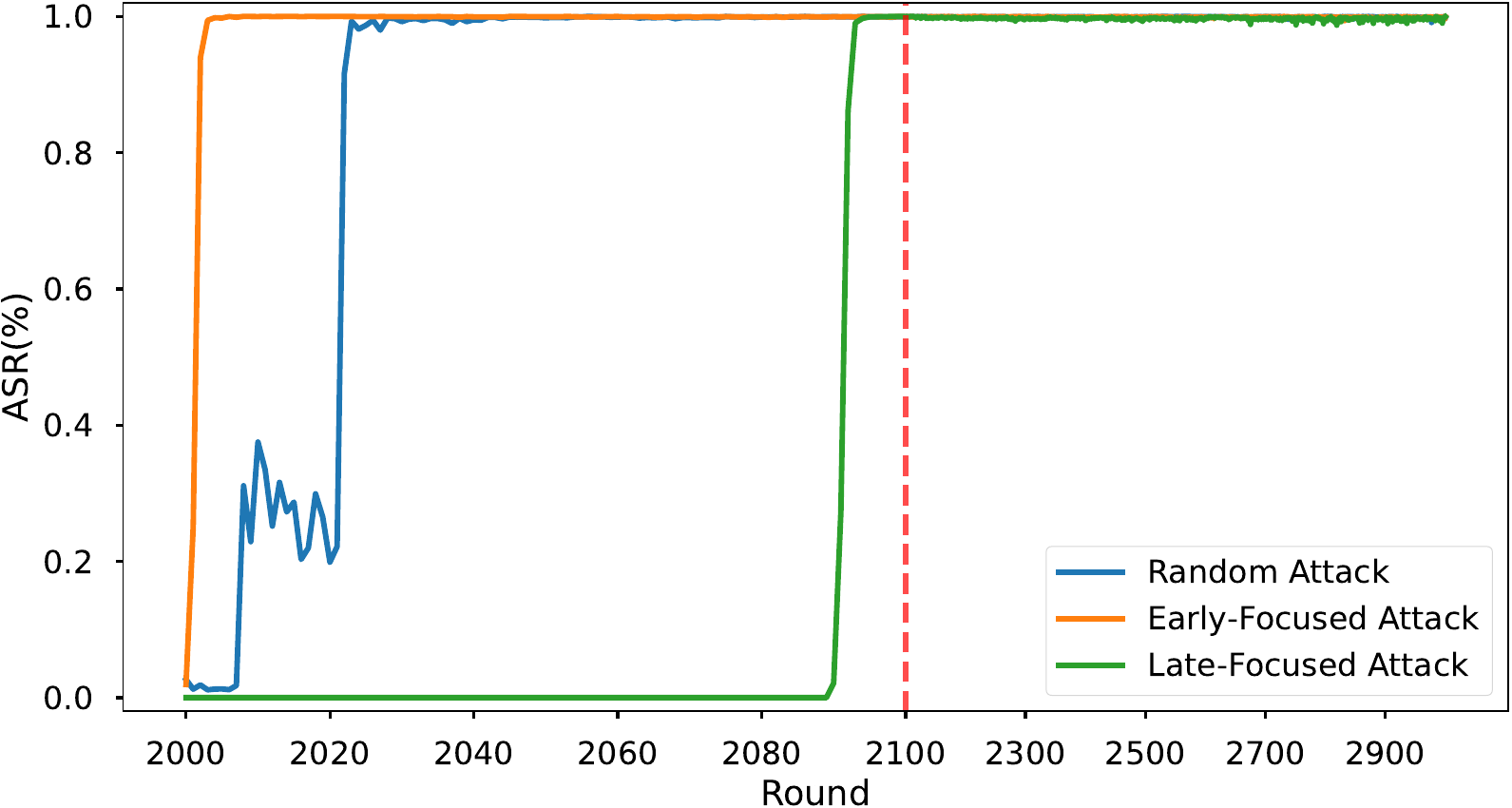}
\caption{Performance of \sysname under different attack time intervals.}
\label{F8}
\vspace{-3mm}
\end{figure}

\textbf{Different Attack Timings.} In this section, we investigate the persistence of \sysname when attacks are executed at different time intervals, demonstrating its applicability in specialized scenarios. Specifically, we configured a single attacker with an attack window spanning rounds 2000-2100, while the FL process continued for a total of 3000 rounds. We established three distinct scenarios: the attacker being randomly selected throughout the attack window, the attacker continuously attacking for 10 consecutive rounds at the beginning of the window (rounds 2000-2010), and the attacker continuously attacking for 10 consecutive rounds at the end of the window (rounds 2090-2100).

As illustrated in Figure \ref{F8}, experimental results reveal remarkable attack effectiveness and persistence regardless of whether the attacks were executed at the beginning or end of the attack window. This temporal robustness can be attributed to \sysname's ability to embed backdoor features in a way that aligns with the model's evolving feature space, ensuring persistent influence even as global training progresses. Unlike conventional approaches that rely on transient model vulnerabilities, our strategy maintains attack efficacy by dynamically adapting to the federated aggregation process without requiring continuous adversarial presence. These findings underscore the practical viability of our method in scenarios where attackers operate intermittently or have limited opportunities to participate in the FL protocol.

In addition, we explore the impact of different trigger types, as well as fixed triggers and optimized triggers on \sysname in Appendix \ref{A-triiger_type}.

\subsection{Parameter Sensitivity and  Ablation}

\begin{figure}
\centering
\includegraphics[width=0.85\linewidth]{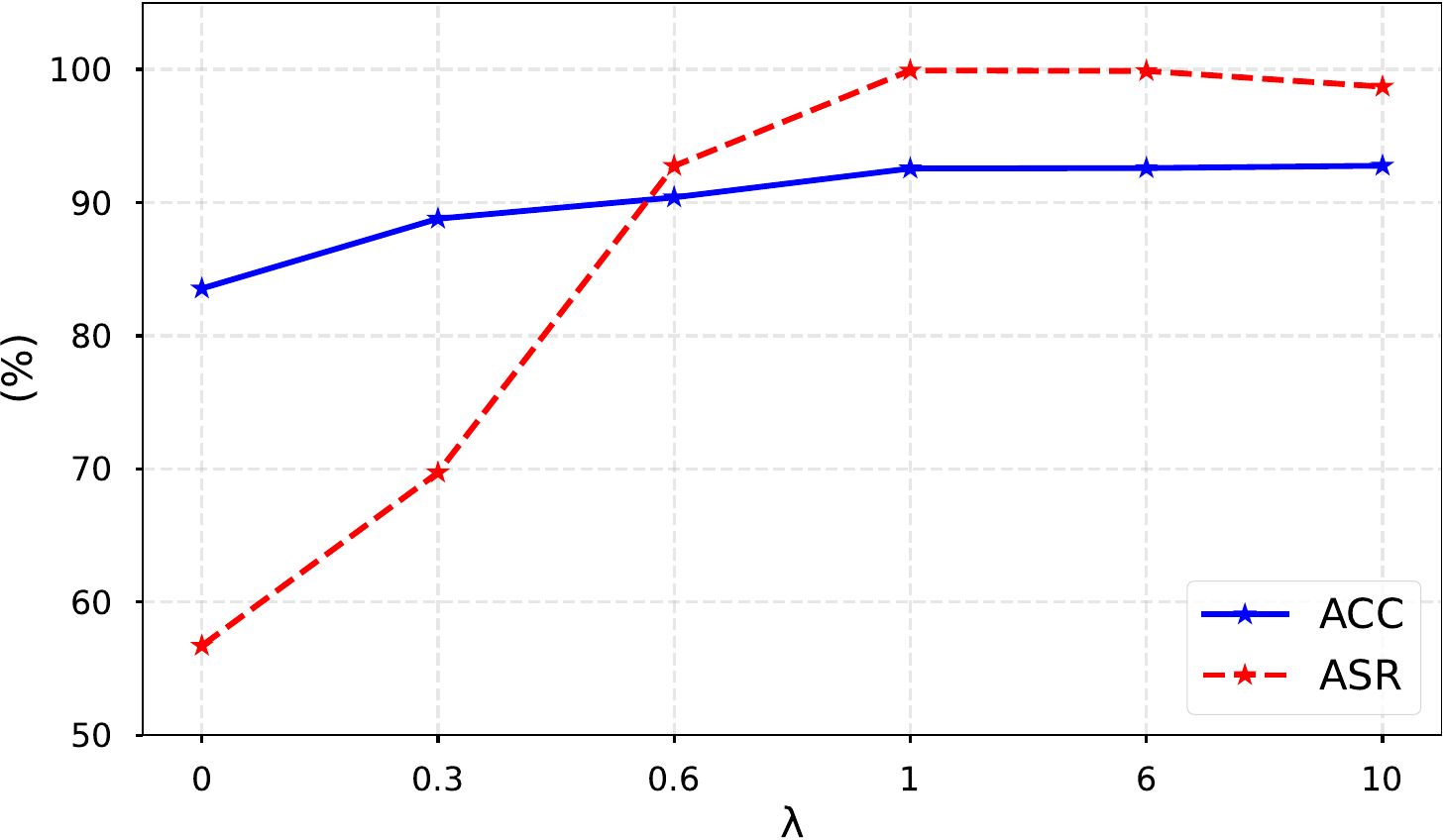}
\caption{Impact of loss term $\lambda$.}
\label{F9}
\vspace{-5mm}
\end{figure}

\textbf{Loss Terms.}
The hyperparameter $\lambda$ holds significant importance in striking a balance between $\mathcal{L}_{align}$ and $\mathcal{L}_{utility}$ during backdoor injection. While the former ensures attack effectiveness through feature alignment, the latter maintains model utility. To investigate the impact of different $\lambda$ values on our method's performance, we varied $\lambda$ across a range of values: 0, 0.3, 0.6, 1, 6, and 10. The experimental results are illustrated in  Figure \ref{F9}.

When $\lambda=0$, indicating the absence of utility constraints, both ASR and model ACC were adversely affected. This observation suggests that the success of the backdoor attack is intricately tied to the overall model performance. As the weight of the utility constraint increases, both ASR and ACC stabilize, suggesting that our feature alignment mechanism operates synergistically with normal model learning rather than competing against it. By aligning backdoor features with legitimate target class features, \sysname creates backdoor pathways that complement rather than contradict the model's primary classification objective. Notably, even at higher values of \(\lambda\), where the utility loss is heavily prioritized, the decline in ASR remains marginal. This limited reduction in attack success rate underscores the resilience of our feature alignment strategy, which maintains high ASR even under stringent utility constraints.

\begin{table}[t]
 \caption{\small Ablation of different components of \sysname.}
  \vspace{1mm}
  \centering
  \footnotesize
  \renewcommand\tabcolsep{2.4pt}
  \renewcommand\arraystretch{1.1}
\begin{tabular}{cccccc}
\toprule
\multicolumn{2}{c}{\textbf{Injection}} & \multicolumn{2}{c}{\textbf{Enhance}} & \multirow{2}{*}{\textbf{ACC}} & \multirow{2}{*}{\textbf{ASR}} \\ \cmidrule{1-4}
$\mathcal{L}_{align}$                     & \small $\mathcal{L}_{utility}$                     & $\mathcal{L}_{enhance}$                    & $\mathcal{L}_{consist}$                    &                               &                               \\ \midrule
\textcolor{dark_green}{\ding{51}}                      &      \textcolor{red}{\ding{55}}                        &      \textcolor{red}{\ding{55}}                       &      \textcolor{red}{\ding{55}}                       & 80.76 (\textcolor{dark_green}{$\downarrow$11.79)}                         & 17.94 (\textcolor{dark_green}{$\downarrow$81.97)}                         \\
\textcolor{dark_green}{\ding{51}}                      &      \textcolor{red}{\ding{55}}                        & \textcolor{dark_green}{\ding{51}}                     &      \textcolor{red}{\ding{55}}                       & 81.85 (\textcolor{dark_green}{$\downarrow$10.70)}                         & 59.52 (\textcolor{dark_green}{$\downarrow$40.39)}                         \\
\textcolor{dark_green}{\ding{51}}                      &      \textcolor{red}{\ding{55}}                        & \textcolor{dark_green}{\ding{51}}                     & \textcolor{dark_green}{\ding{51}}                     & 83.54 (\textcolor{dark_green}{$\downarrow$9.01)}                         & 56.72 (\textcolor{dark_green}{$\downarrow$43.19)}                        \\
\textcolor{dark_green}{\ding{51}}                      & \textcolor{dark_green}{\ding{51}}                      &      \textcolor{red}{\ding{55}}                       &      \textcolor{red}{\ding{55}}                       & 89.22 (\textcolor{dark_green}{$\downarrow$3.33)}                         & 68.05 (\textcolor{dark_green}{$\downarrow$31.86)}                         \\
\textcolor{dark_green}{\ding{51}}                      & \textcolor{dark_green}{\ding{51}}                      & \textcolor{dark_green}{\ding{51}}                     &      \textcolor{red}{\ding{55}}                       & 91.04 (\textcolor{dark_green}{$\downarrow$1.54)}                        & 94.37 (\textcolor{dark_green}{$\downarrow$5.54)}                         \\
\textcolor{dark_green}{\ding{51}}                      & \textcolor{dark_green}{\ding{51}}                      & \textcolor{dark_green}{\ding{51}}                     & \textcolor{dark_green}{\ding{51}}                     & \textbf{92.55}                & \textbf{99.91}                \\ \bottomrule
\end{tabular}
  \label{T4}
 \vspace{-3mm}
\end{table}

\begin{table}[t]
 \caption{\small Performance of \sysname with different constraint norms.}
  \vspace{1mm}
  \centering
  \footnotesize
  \renewcommand\tabcolsep{10pt}
  \renewcommand\arraystretch{1.2}
\begin{tabular}{ccc}
\toprule
\textbf{Constraint Method}  & \textbf{ACC (\%)} & \textbf{ASR (\%)} \\ \midrule
\textbf{Lp-Norms}           & 91.51 (\textcolor{dark_green}{$\downarrow$1.04)}            & 88.79 (\textcolor{dark_green}{$\downarrow$11.12)}            \\
\textbf{KL Divergence}      & 87.73 (\textcolor{dark_green}{$\downarrow$4.82)}            & 69.37 (\textcolor{dark_green}{$\downarrow$30.54)}            \\
\textbf{Cosine Similarity}  & 92.12 (\textcolor{dark_green}{$\downarrow$0.43)}            & 94.58 (\textcolor{dark_green}{$\downarrow$5.33)}            \\ \midrule
\textbf{Sliced-Wasserstein} & \textbf{92.55}    & \textbf{99.91}    \\ \bottomrule
\end{tabular}
  \label{T5}
\end{table}

\textbf{Component Contributions.}
\sysname consists of two integral components: backdoor injection and backdoor enhancement. In the backdoor injection phase, we employ the $\mathcal{L}_{align}$ loss to ensure attack effectiveness, while simultaneously utilizing $\mathcal{L}_{utility}$ loss to preserve model utility. The backdoor enhancement component leverages $\mathcal{L}_{enhance}$ loss to amplify trigger effectiveness, while maintaining trigger stealthiness through feature consistency loss $\mathcal{L}_{consist}$. In this section, we further validate the contribution of each loss term to the overall framework. Since $\mathcal{L}_{align}$ represents the fundamental mechanism for backdoor injection, we retain it in all experimental configurations.

The experimental results presented in Table \ref{T4} demonstrate that each loss term plays a crucial role in our attack framework.  When removing $\mathcal{L}_{utility}$, we observe a significant degradation in the model's primary task performance. This confirms our hypothesis that preserving model utility is essential for successful backdoor injection in our framework, as our approach fundamentally relies on aligning trigger features with legitimate target class features. Without maintaining model utility, these target class features become distorted, making it difficult to establish the necessary feature connections for effective backdoor functionality. Moreover, while $\mathcal{L}_{enhance}$ significantly boosts ASR through adversarial trigger optimization, its effectiveness heavily relies on the concurrent application of $\mathcal{L}_{consist}$ to maintain feature-space consistency. Without this constraint, the enhanced triggers tend to deviate from the target distribution, becoming more susceptible to state-of-the-art anomaly detection methods. The feature projection distance metric in $\mathcal{L}_{consist}$ proves particularly effective in balancing attack potency with stealth, as it enforces directional alignment rather than absolute magnitude matching, making the triggers more resilient against model updates during federated aggregation rounds.

\begin{figure}[t]
\centering
\includegraphics[width=0.90\linewidth]{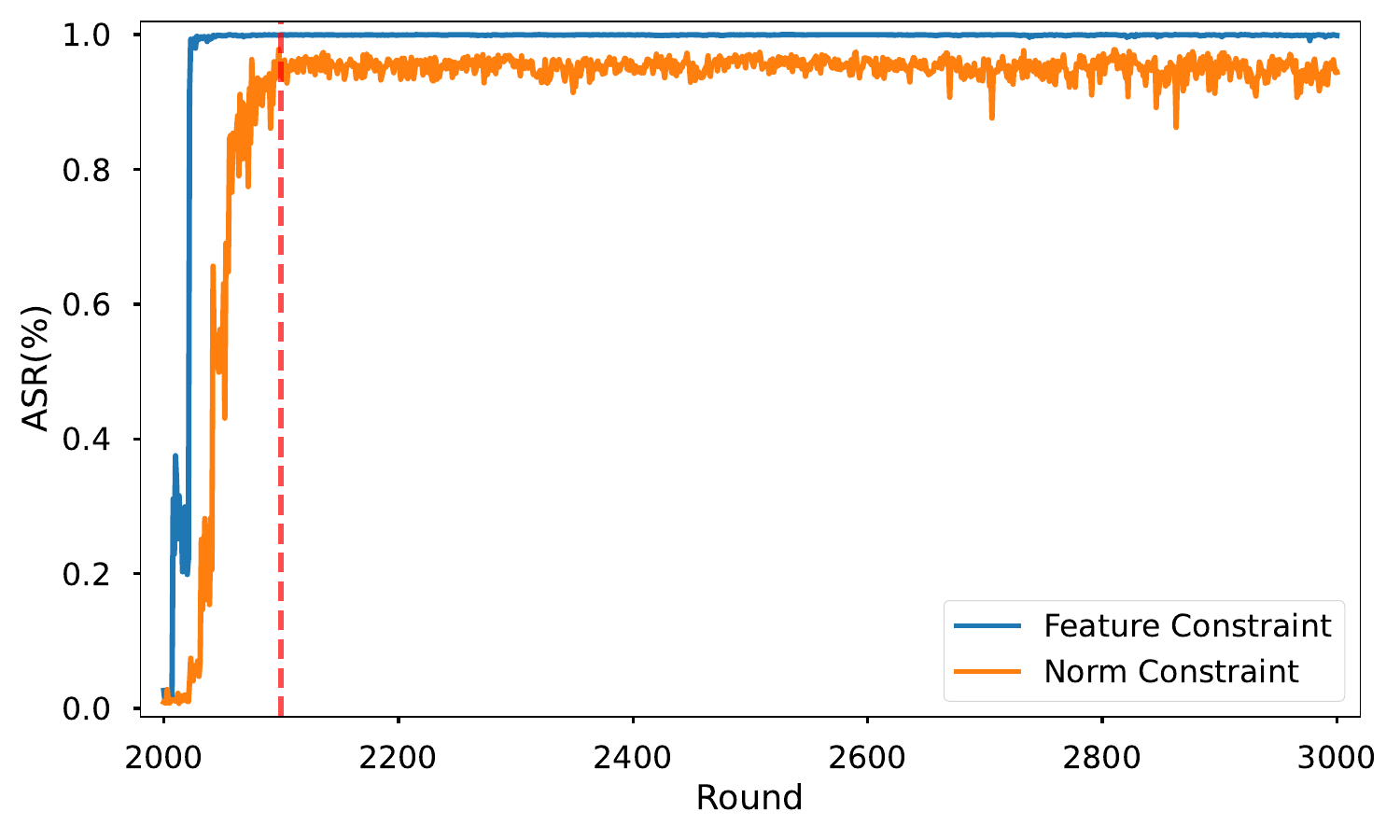}
\caption{Performance of \sysname using different trigger constraints.}
\label{F10}
\vspace{-3mm}
\end{figure}

\begin{figure}[t]
\centering
\includegraphics[width=1\linewidth]{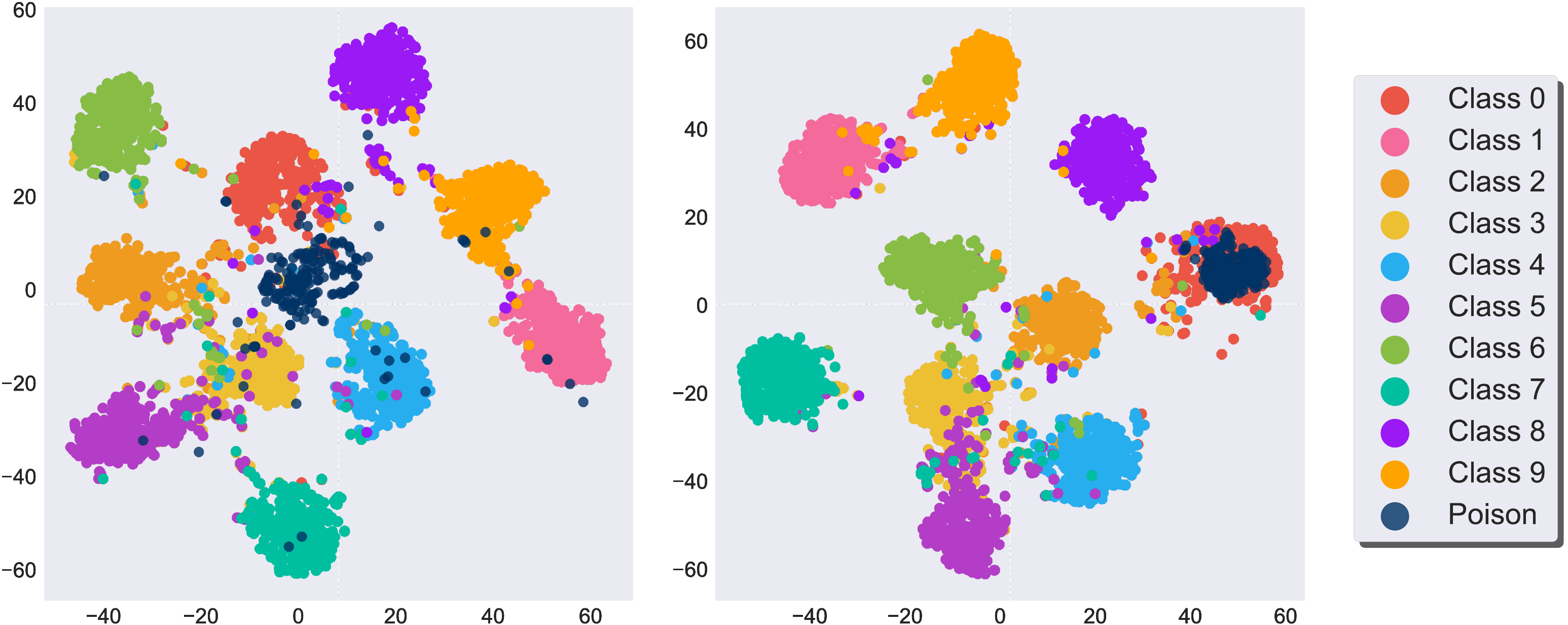}
\caption{T-SNE visualization of \sysname using different trigger constraints.}
\label{F11}
\vspace{-3mm}
\end{figure}

\textbf{Different Constraint Norms.}
This section investigates the impact of different distance metrics on backdoor injection performance during the feature alignment phase. We evaluate three commonly used metrics: $L_2$ norm, cosine similarity, and KL divergence, with comparative results presented in Table \ref{T5}. Consistent with prior research, our experiments reveal that $L_2$ norm's point-wise alignment constraint tends to distort critical feature channels, which not only degrades the model's primary classification accuracy but also inadvertently limits backdoor effectiveness. While cosine similarity better preserves the main task performance by focusing on directional alignment, it imposes overly strict constraints that similarly hinder optimal backdoor injection.  The KL divergence exhibited the most pronounced drawbacks, with both ACC and ASR experiencing significant declines. This underperformance can be attributed to the numerical instability of KL divergence in high-dimensional spaces and its reliance on accurate density estimation, which is particularly challenging in the dynamic and heterogeneous feature spaces characteristic of FL. 


In contrast, \sysname leverages the Sliced Wasserstein Distance (SWD), which is specifically designed to address the challenges of high-dimensional distribution matching. The SWD excels at capturing the dynamic variations in the feature space of FL models by approximating the optimal transport between distributions through one-dimensional projections. This approach ensures a more robust and flexible alignment of backdoor features with the target class’s distribution, mitigating the distortions observed with the $L_2$ norm and the limitations of cosine similarity and KL divergence.

\textbf{Trigger Constraint.}
In this section, we evaluate the performance of two distinct approaches for constraining trigger magnitude: direct L-norm restrictions and feature consistency constraints. Additionally, we visualize the T-SNE projections for both scenarios. The experimental results are presented in Table \ref{F10}. While both methods demonstrate effectiveness, we observe that direct L-norm restrictions impose certain limitations on the ASR. 

More significantly, the T-SNE visualizations shown in Figure \ref{F11} reveal a critical distinction in feature distribution patterns. Trigger features constrained by L-norm restrictions predominantly cluster at the periphery of the target class feature space, creating a discernible boundary that potentially increases detection risk. In contrast, trigger features constrained through our feature alignment approach exhibit distributions that more closely resemble normal features, positioned well within the legitimate feature clusters of the target class. This deeper integration with legitimate feature representations substantially enhances the attack's stealthiness and resistance to detection.

\section{Summary and Future Work}\label{Sec: con}
In this paper, we introduced \sysname, a novel backdoor attack framework targeting federated learning systems. Departing from conventional end-to-end supervised backdoor injection, \sysname instead aligns feature representations of trigger-embedded samples with those of the target class, greatly enhancing stealth and long-term persistence. Our adversarial trigger optimization mechanism further improves the attack’s adaptability and effectiveness, allowing backdoors to survive sophisticated aggregation schemes and persist long after adversarial participation ceases.  Our comprehensive empirical evaluation demonstrates that \sysname significantly outperforms state-of-the-art backdoor attacks across various metrics and scenarios. The attack maintains high success rates and remarkable persistence even under challenging conditions such as highly skewed data distributions and limited adversary participation. Additionally, \sysname shows robust performance across different trigger implementations and successfully targets multiple labels simultaneously without compromising attack effectiveness. This highlights the urgent demand for tailored defenses. In future work, we will explore feature-level backdoor defense methods to target complex dynamic backdoor attacks in FL.


\bibliographystyle{IEEEtran}
\bibliography{main}

\begin{thebibliography}{10}
\providecommand{\url}[1]{#1}
\csname url@samestyle\endcsname
\providecommand{\newblock}{\relax}
\providecommand{\bibinfo}[2]{#2}
\providecommand{\BIBentrySTDinterwordspacing}{\spaceskip=0pt\relax}
\providecommand{\BIBentryALTinterwordstretchfactor}{4}
\providecommand{\BIBentryALTinterwordspacing}{\spaceskip=\fontdimen2\font plus
\BIBentryALTinterwordstretchfactor\fontdimen3\font minus \fontdimen4\font\relax}
\providecommand{\BIBforeignlanguage}[2]{{%
\expandafter\ifx\csname l@#1\endcsname\relax
\typeout{** WARNING: IEEEtran.bst: No hyphenation pattern has been}%
\typeout{** loaded for the language `#1'. Using the pattern for}%
\typeout{** the default language instead.}%
\else
\language=\csname l@#1\endcsname
\fi
#2}}
\providecommand{\BIBdecl}{\relax}
\BIBdecl

\bibitem{FL_mcmahan_2017}
B.~McMahan, E.~Moore, D.~Ramage, S.~Hampson, and B.~A. y~Arcas, ``{{Communication-Efficient Learning}} of {{Deep Networks}} from {{Decentralized Data}},'' in \emph{Proceedings of the 20th {{International Conference}} on {{Artificial Intelligence}} and {{Statistics}}}.\hskip 1em plus 0.5em minus 0.4em\relax PMLR, Apr. 2017, pp. 1273--1282.

\bibitem{HeteroFair_Li_2024}
Y.~Li, J.~Zhang, Y.~Zhao, B.~Chen, and S.~Yu, ``Fairness-aware federated learning framework on heterogeneous data distributions,'' in \emph{ICC 2024-IEEE International Conference on Communications}.\hskip 1em plus 0.5em minus 0.4em\relax IEEE, 2024, pp. 728--733.

\bibitem{GBoard_hard_2018}
A.~Hard, K.~Rao, R.~Mathews, S.~Ramaswamy, F.~Beaufays, S.~Augenstein, H.~Eichner, C.~Kiddon, and D.~Ramage, ``{{Federated Learning}} for {{Mobile Keyboard Prediction}},'' Nov. 2018.

\bibitem{QuickType_Apple_2019}
Apple, ``Private federated learning (neurips 2019 expo talk abstract),'' \url{https://nips.cc/ExpoConferences/2019/schedule?talk id=40}, accessed: 2020-05-22.

\bibitem{Financial_long_2020}
G.~Long, Y.~Tan, J.~Jiang, and C.~Zhang, ``Federated learning for open banking,'' in \emph{Federated learning: privacy and incentive}.\hskip 1em plus 0.5em minus 0.4em\relax Springer, 2020, pp. 240--254.

\bibitem{Medical_sheller_2019}
M.~J. Sheller, G.~A. Reina, B.~Edwards, J.~Martin, and S.~Bakas, ``{{Multi-institutional Deep Learning Modeling Without Sharing Patient Data}}: {{A Feasibility Study}} on {{Brain Tumor Segmentation}},'' in \emph{Brainlesion: {{Glioma}}, {{Multiple Sclerosis}}, {{Stroke}} and {{Traumatic Brain Injuries}}}.\hskip 1em plus 0.5em minus 0.4em\relax Cham: Springer International Publishing, 2019, pp. 92--104.

\bibitem{security}
Y.~Liu, Y.~Kang, T.~Zou, Y.~Pu, Y.~He, X.~Ye, Y.~Ouyang, Y.-Q. Zhang, and Q.~Yang, ``Vertical federated learning: Concepts, advances, and challenges,'' \emph{IEEE Transactions on Knowledge \& Data Engineering}, vol.~36, no.~07, pp. 3615--3634, 2024.

\bibitem{feng2023infer1}
C.-M. Feng, K.~Yu, N.~Liu, X.~Xu, S.~Khan, and W.~Zuo, ``Towards instance-adaptive inference for federated learning,'' in \emph{Proceedings of the IEEE/CVF International Conference on Computer Vision}, 2023, pp. 23\,287--23\,296.

\bibitem{fu2022infer2}
C.~Fu, X.~Zhang, S.~Ji, J.~Chen, J.~Wu, S.~Guo, J.~Zhou, A.~X. Liu, and T.~Wang, ``Label inference attacks against vertical federated learning,'' in \emph{31st USENIX security symposium (USENIX Security 22)}, 2022, pp. 1397--1414.

\bibitem{ma2022posion1}
Z.~Ma, J.~Ma, Y.~Miao, Y.~Li, and R.~H. Deng, ``Shieldfl: Mitigating model poisoning attacks in privacy-preserving federated learning,'' \emph{IEEE Transactions on Information Forensics and Security}, vol.~17, pp. 1639--1654, 2022.

\bibitem{fang2020posion2}
M.~Fang, X.~Cao, J.~Jia, and N.~Gong, ``Local model poisoning attacks to $\{$Byzantine-Robust$\}$ federated learning,'' in \emph{29th USENIX security symposium (USENIX Security 20)}, 2020, pp. 1605--1622.

\bibitem{A3FL_Zhang_2024}
H.~Zhang, J.~Jia, J.~Chen, L.~Lin, and D.~Wu, ``{A3FL}: adversarially adaptive backdoor attacks to federated learning,'' in \emph{Proceedings of the 37th International Conference on Neural Information Processing Systems}, 2023, pp. 61\,213--61\,233.

\bibitem{backdoorFL_bagdasaryan_2020}
E.~Bagdasaryan, A.~Veit, Y.~Hua, D.~Estrin, and V.~Shmatikov, ``How to backdoor federated learning,'' in \emph{International conference on artificial intelligence and statistics}.\hskip 1em plus 0.5em minus 0.4em\relax PMLR, 2020, pp. 2938--2948.

\bibitem{Neurotoxin_zhang_2022}
Z.~Zhang, A.~Panda, L.~Song, Y.~Yang, M.~Mahoney, P.~Mittal, R.~Kannan, and J.~Gonzalez, ``Neurotoxin: Durable backdoors in federated learning,'' in \emph{International Conference on Machine Learning}.\hskip 1em plus 0.5em minus 0.4em\relax PMLR, 2022, pp. 26\,429--26\,446.

\bibitem{Chameleon_dai_2023}
Y.~Dai and S.~Li, ``Chameleon: Adapting to peer images for planting durable backdoors in federated learning,'' in \emph{International Conference on Machine Learning}.\hskip 1em plus 0.5em minus 0.4em\relax PMLR, 2023, pp. 6712--6725.

\bibitem{Advdoor_zhang_2021}
Q.~Zhang, Y.~Ding, Y.~Tian, J.~Guo, M.~Yuan, and Y.~Jiang, ``Advdoor: adversarial backdoor attack of deep learning system,'' in \emph{Proceedings of the 30th ACM SIGSOFT International Symposium on Software Testing and Analysis}, 2021, pp. 127--138.

\bibitem{Darkfed_li_2024}
M.~Li, W.~Wan, Y.~Ning, S.~Hu, L.~Xue, L.~Y. Zhang, and Y.~Wang, ``Darkfed: A data-free backdoor attack in federated learning,'' \emph{arXiv preprint arXiv:2405.03299}, 2024.

\bibitem{li2021anti}
Y.~Li, X.~Lyu, N.~Koren, L.~Lyu, B.~Li, and X.~Ma, ``Anti-backdoor learning: Training clean models on poisoned data,'' \emph{Advances in Neural Information Processing Systems}, vol.~34, pp. 14\,900--14\,912, 2021.

\bibitem{FLAME_Nguyen_2022}
T.~D. Nguyen, P.~Rieger, H.~Chen, H.~Yalame, H.~M{\"{o}}llering, H.~Fereidooni, S.~Marchal, M.~Miettinen, A.~Mirhoseini, S.~Zeitouni, F.~Koushanfar, A.~Sadeghi, and T.~Schneider, ``{FLAME:} taming backdoors in federated learning,'' in \emph{31st {USENIX} Security Symposium, {USENIX} Security 2022, Boston, MA, USA, August 10-12, 2022}.\hskip 1em plus 0.5em minus 0.4em\relax {USENIX} Association, 2022, pp. 1415--1432.

\bibitem{khosla2020contrastive}
P.~Khosla, P.~Teterwak, C.~Wang, A.~Sarna, Y.~Tian, P.~Isola, A.~Maschinot, C.~Liu, and D.~Krishnan, ``Supervised contrastive learning,'' \emph{Advances in neural information processing systems}, vol.~33, pp. 18\,661--18\,673, 2020.

\bibitem{BadNets_gu_2019}
T.~Gu, K.~Liu, B.~{Dolan-Gavitt}, and S.~Garg, ``{{BadNets}}: {{Evaluating Backdooring Attacks}} on {{Deep Neural Networks}},'' \emph{IEEE Access}, vol.~7, pp. 47\,230--47\,244, 2019.

\bibitem{Blend_chen_2017}
X.~Chen, C.~Liu, B.~Li, K.~Lu, and D.~Song, ``Targeted backdoor attacks on deep learning systems using data poisoning,'' \emph{arXiv preprint arXiv:1712.05526}, 2017.

\bibitem{wang2020attack}
H.~Wang, K.~Sreenivasan, S.~Rajput, H.~Vishwakarma, S.~Agarwal, J.-y. Sohn, K.~Lee, and D.~Papailiopoulos, ``Attack of the tails: Yes, you really can backdoor federated learning,'' \emph{Advances in Neural Information Processing Systems}, vol.~33, pp. 16\,070--16\,084, 2020.

\bibitem{DBA_Xie_2020}
C.~Xie, K.~Huang, P.~Chen, and B.~Li, ``{DBA:} distributed backdoor attacks against federated learning,'' in \emph{8th International Conference on Learning Representations, {ICLR} 2020, Addis Ababa, Ethiopia, April 26-30, 2020}.\hskip 1em plus 0.5em minus 0.4em\relax OpenReview.net, 2020.

\bibitem{Towards_Shi_2024}
C.~Shi, S.~Ji, X.~Pan, X.~Zhang, M.~Zhang, M.~Yang, J.~Zhou, J.~Yin, and T.~Wang, ``Towards practical backdoor attacks on federated learning systems,'' \emph{IEEE Transactions on Dependable and Secure Computing}, 2024.

\bibitem{F3BA_Fang_2023}
P.~Fang and J.~Chen, ``On the vulnerability of backdoor defenses for federated learning,'' in \emph{Proceedings of the AAAI Conference on Artificial Intelligence}, vol.~37, no.~10, 2023, pp. 11\,800--11\,808.

\bibitem{3DFed_Li_2023}
H.~Li, Q.~Ye, H.~Hu, J.~Li, L.~Wang, C.~Fang, and J.~Shi, ``3dfed: Adaptive and extensible framework for covert backdoor attack in federated learning,'' in \emph{44th {IEEE} Symposium on Security and Privacy, {SP} 2023, San Francisco, CA, USA, May 21-25, 2023}.\hskip 1em plus 0.5em minus 0.4em\relax {IEEE}, 2023, pp. 1893--1907.

\bibitem{CerP_Lyu_2023}
X.~Lyu, Y.~Han, W.~Wang, J.~Liu, B.~Wang, J.~Liu, and X.~Zhang, ``Poisoning with cerberus: Stealthy and colluded backdoor attack against federated learning,'' in \emph{Proceedings of the AAAI Conference on Artificial Intelligence}, vol.~37, no.~7, 2023, pp. 9020--9028.

\bibitem{multikrum_Peva_2017}
P.~Blanchard, E.~M. El~Mhamdi, R.~Guerraoui, and J.~Stainer, ``Machine learning with adversaries: Byzantine tolerant gradient descent,'' in \emph{Advances in Neural Information Processing Systems}, vol.~30.\hskip 1em plus 0.5em minus 0.4em\relax Curran Associates, Inc., 2017.

\bibitem{Foolsgold_Fung_2022}
C.~Fung, C.~J.~M. Yoon, and I.~Beschastnikh, ``The limitations of federated learning in sybil settings,'' in \emph{23rd International Symposium on Research in Attacks, Intrusions and Defenses, {RAID} 2020, San Sebastian, Spain, October 14-15, 2020}.\hskip 1em plus 0.5em minus 0.4em\relax {USENIX} Association, 2020, pp. 301--316.

\bibitem{RFLBAT_Wang_2022}
\BIBentryALTinterwordspacing
Y.~Wang, D.~Zhai, Y.~Zhan, and Y.~Xia, ``{RFLBAT:} {A} robust federated learning algorithm against backdoor attack,'' \emph{CoRR}, vol. abs/2201.03772, 2022. [Online]. Available: \url{https://arxiv.org/abs/2201.03772}
\BIBentrySTDinterwordspacing

\bibitem{DeepSight_Rieger_2022}
P.~Rieger, T.~D. Nguyen, M.~Miettinen, and A.~Sadeghi, ``Deepsight: Mitigating backdoor attacks in federated learning through deep model inspection,'' in \emph{29th Annual Network and Distributed System Security Symposium, {NDSS} 2022, San Diego, California, USA, April 24-28, 2022}.\hskip 1em plus 0.5em minus 0.4em\relax The Internet Society, 2022.

\bibitem{Backdoorindicator_Li_2024}
S.~Li and Y.~Dai, ``{BackdoorIndicator}: Leveraging {OOD} data for proactive backdoor detection in federated learning,'' in \emph{33rd USENIX Security Symposium (USENIX Security 24)}.\hskip 1em plus 0.5em minus 0.4em\relax Philadelphia, PA: USENIX Association, Aug. 2024, pp. 4193--4210.

\bibitem{xie2021crfl}
C.~Xie, M.~Chen, P.-Y. Chen, and B.~Li, ``Crfl: Certifiably robust federated learning against backdoor attacks,'' in \emph{International Conference on Machine Learning}.\hskip 1em plus 0.5em minus 0.4em\relax PMLR, 2021, pp. 11\,372--11\,382.

\bibitem{RLR2021}
M.~S. Ozdayi, M.~Kantarcioglu, and Y.~R. Gel, ``Defending against backdoors in federated learning with robust learning rate,'' in \emph{Proceedings of the AAAI Conference on Artificial Intelligence}, vol.~35, no.~10, 2021, pp. 9268--9276.

\bibitem{PGD_Wang_2020}
H.~Wang, K.~Sreenivasan, S.~Rajput, H.~Vishwakarma, S.~Agarwal, J.~Sohn, K.~Lee, and D.~S. Papailiopoulos, ``Attack of the tails: Yes, you really can backdoor federated learning,'' in \emph{Advances in Neural Information Processing Systems 33: Annual Conference on Neural Information Processing Systems 2020, NeurIPS 2020, December 6-12, 2020, virtual}, 2020.

\bibitem{DRUPE_tao_2024}
G.~Tao, Z.~Wang, S.~Feng, G.~Shen, S.~Ma, and X.~Zhang, ``{{DRUPE}}: {{Distribution Preserving Backdoor Attack}} in {{Self-supervised Learning}},'' in \emph{2024 IEEE Symposium on Security and Privacy (SP)}.\hskip 1em plus 0.5em minus 0.4em\relax IEEE Computer Society, 2023, pp. 29--29.

\bibitem{WSBA}
\BIBentryALTinterwordspacing
K.~Doan, Y.~Lao, and P.~Li, ``Backdoor attack with imperceptible input and latent modification,'' in \emph{Advances in Neural Information Processing Systems}, M.~Ranzato, A.~Beygelzimer, Y.~Dauphin, P.~Liang, and J.~W. Vaughan, Eds., vol.~34.\hskip 1em plus 0.5em minus 0.4em\relax Curran Associates, Inc., 2021, pp. 18\,944--18\,957. [Online]. Available: \url{https://proceedings.neurips.cc/paper_files/paper/2021/file/9d99197e2ebf03fc388d09f1e94af89b-Paper.pdf}
\BIBentrySTDinterwordspacing

\bibitem{Cifar_Krizhevsky_2009}
A.~Krizhevsky, G.~Hinton \emph{et~al.}, ``Learning multiple layers of features from tiny images.''\hskip 1em plus 0.5em minus 0.4em\relax Toronto, ON, Canada, 2009.

\bibitem{GTSRB_houben_2013}
S.~Houben, J.~Stallkamp, J.~Salmen, M.~Schlipsing, and C.~Igel, ``Detection of traffic signs in real-world images: The german traffic sign detection benchmark,'' in \emph{The 2013 international joint conference on neural networks (IJCNN)}.\hskip 1em plus 0.5em minus 0.4em\relax IEEE, 2013, pp. 1--8.

\bibitem{Resnet_He_2016}
\BIBentryALTinterwordspacing
K.~He, X.~Zhang, S.~Ren, and J.~Sun, ``Deep residual learning for image recognition,'' in \emph{2016 {IEEE} Conference on Computer Vision and Pattern Recognition, {CVPR} 2016, Las Vegas, NV, USA, June 27-30, 2016}.\hskip 1em plus 0.5em minus 0.4em\relax {IEEE} Computer Society, 2016, pp. 770--778. [Online]. Available: \url{https://doi.org/10.1109/CVPR.2016.90}
\BIBentrySTDinterwordspacing

\bibitem{VGG_Simonyan_2015}
K.~Simonyan and A.~Zisserman, ``Very deep convolutional networks for large-scale image recognition,'' in \emph{3rd International Conference on Learning Representations, {ICLR} 2015, San Diego, CA, USA, May 7-9, 2015, Conference Track Proceedings}, 2015.

\bibitem{Mobilenetv2_sandler_2018}
M.~Sandler, A.~Howard, M.~Zhu, A.~Zhmoginov, and L.-C. Chen, ``Mobilenetv2: Inverted residuals and linear bottlenecks,'' in \emph{Proceedings of the IEEE conference on computer vision and pattern recognition}, 2018, pp. 4510--4520.

\bibitem{zhu2021data}
Z.~Zhu, J.~Hong, and J.~Zhou, ``Data-free knowledge distillation for heterogeneous federated learning,'' in \emph{International Conference on Machine Learning}.\hskip 1em plus 0.5em minus 0.4em\relax PMLR, 2021, pp. 12\,878--12\,889.

\bibitem{BadCleaner_JLZ_2024}
J.~Zhang, C.~Zhu, C.~Ge, C.~Ma, Y.~Zhao, X.~Sun, and B.~Chen, ``{{BadCleaner}}: {{Defending Backdoor Attacks}} in {{Federated Learning}} via {{Attention-Based Multi-Teacher Distillation}},'' \emph{IEEE Transactions on Dependable and Secure Computing}, vol.~21, no.~5, pp. 4559--4573, Sep. 2024.

\bibitem{FLPurifier_JLZ_2024}
J.~Zhang, C.~Zhu, X.~Sun, C.~Ge, B.~Chen, W.~Susilo, and S.~Yu, ``{{{\emph{FLPurifier}}}} : {{Backdoor Defense}} in {{Federated Learning}} via {{Decoupled Contrastive Training}},'' \emph{IEEE Transactions on Information Forensics and Security}, vol.~19, pp. 4752--4766, 2024.

\bibitem{Dirichlet_hsu_2019}
T.-M.~H. Hsu, H.~Qi, and M.~Brown, ``Measuring the effects of non-identical data distribution for federated visual classification,'' \emph{arXiv preprint arXiv:1909.06335}, 2019.

\end{thebibliography}

\appendices

\begin{figure*}[h]
	\centering
	\subfigure[Results on CIFAR-10 dataset.]{
		\begin{minipage}[t]{0.322\linewidth}
			\centering
			\includegraphics[width=1\linewidth]{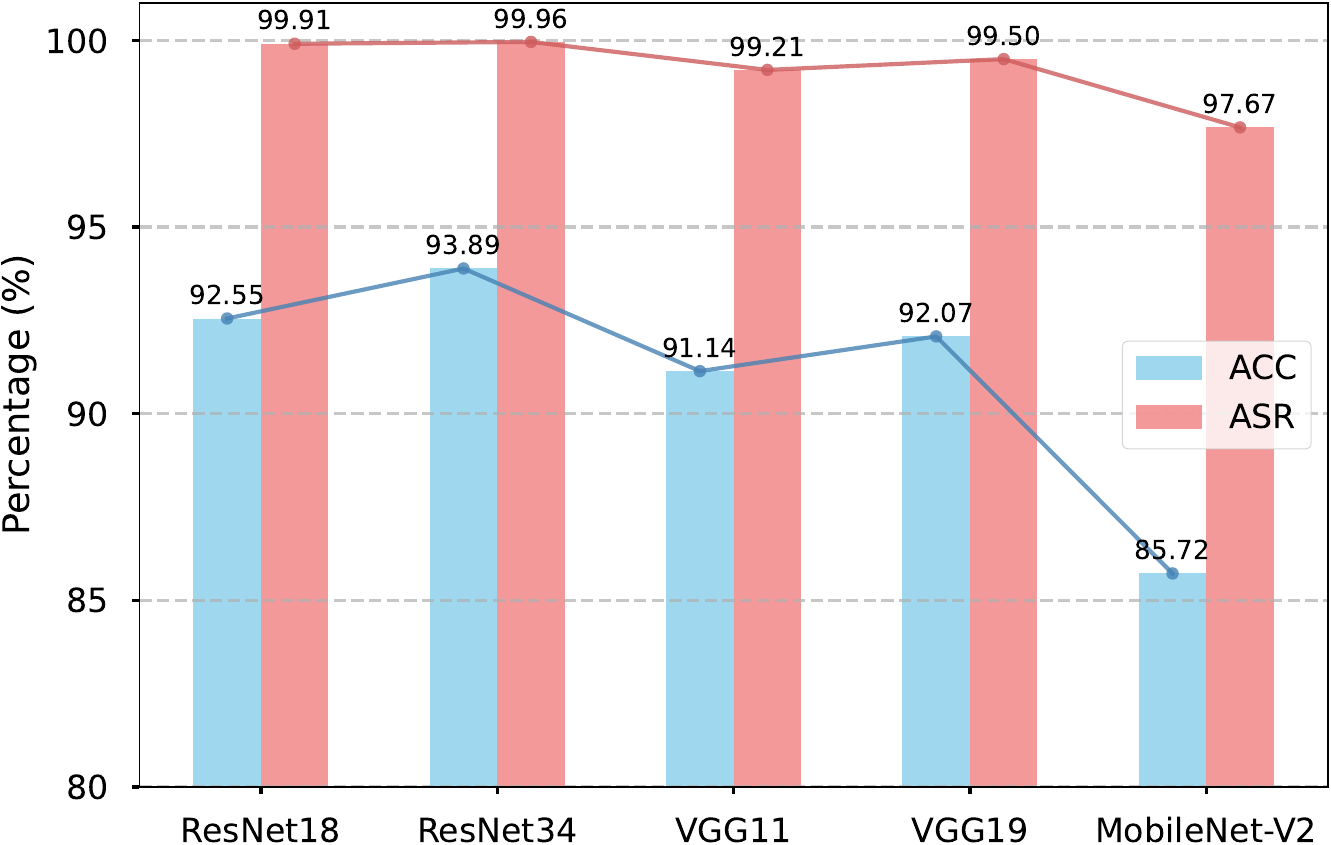}
			\label{F12a}
	\end{minipage}}
	\subfigure[Results on CIFAR-100 dataset.]{
		\begin{minipage}[t]{0.322\linewidth}
			\includegraphics[width=1\linewidth]{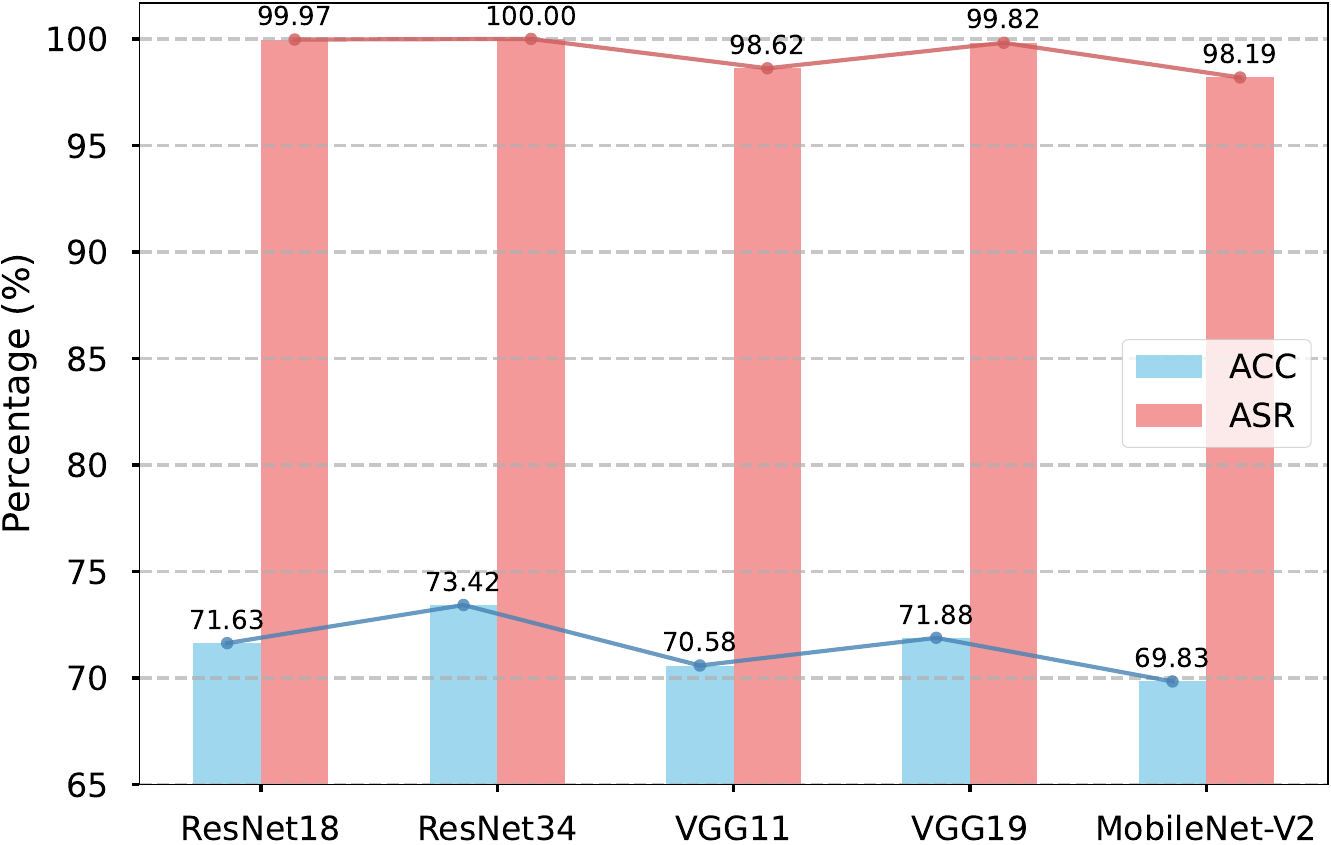}
			\label{F12b}
	\end{minipage}}
	\subfigure[Results on GTSRB dataset.]{
		\begin{minipage}[t]{0.322\linewidth}
			\centering
			\includegraphics[width=1\linewidth]{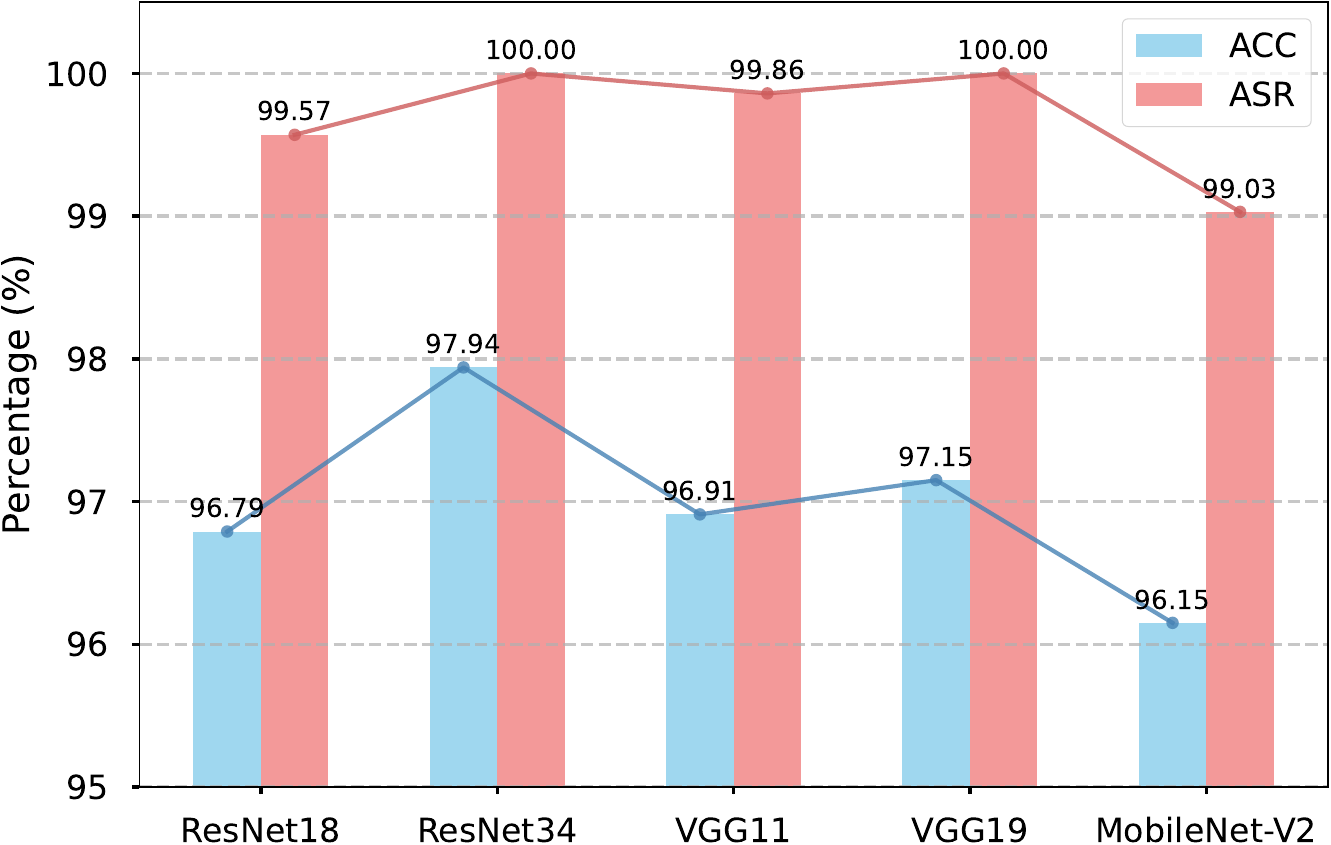}
			\label{F12c}
	\end{minipage}}
	\caption{Impact of model types on three datasets.}
	\label{F12}
	\vspace{-2mm}
\end{figure*}

\section{Impact of Different Model Architectures} 
\label{A-Model_type}
In the context of backdoor attacks, the prevailing approach typically involves manipulating training data to embed malicious patterns, enabling the model to learn a spurious correlation between these tampered inputs and specific target labels through supervised training. This strategy is generally agnostic to the internal architecture of the model, focusing solely on the input-output relationship without regard for the model’s structural nuances. In contrast, our proposed method diverges by establishing a backdoor attack through the alignment of backdoor features with the feature representations of the target class in the feature space. Given that different model architectures inherently learn distinct feature representations, it becomes imperative to evaluate the robustness of our method across varying model structures to ascertain its generalizability. We conduct comprehensive experiments across six widely used image classification architectures: ResNet18, ResNet34, VGG11, VGG19, and MobileNet-V2, while keeping all other parameters consistent with our default configuration.

The results, presented in Figure \ref{F12}, reveal several important insights about architectural sensitivity. While the ACC varies across different models, the ASR remains consistently high (above 98$\%$) regardless of architecture. Interestingly, we observed a positive correlation between ACC and ASR to a certain degree. Based on our analysis, this correlation exists because higher model accuracy generally indicates stronger feature representation capabilities, resulting in better target class representation. When backdoor features align with these well-defined target class features, they achieve higher attack success rates. Models with superior representation learning abilities provide clearer feature boundaries and more distinct class characteristics, which our method can leverage more effectively for backdoor feature alignment.

\section{Feature Space Visualization}
\label{A_visual}

\begin{figure*}[h]
	\centering
	\includegraphics[width=1\linewidth]{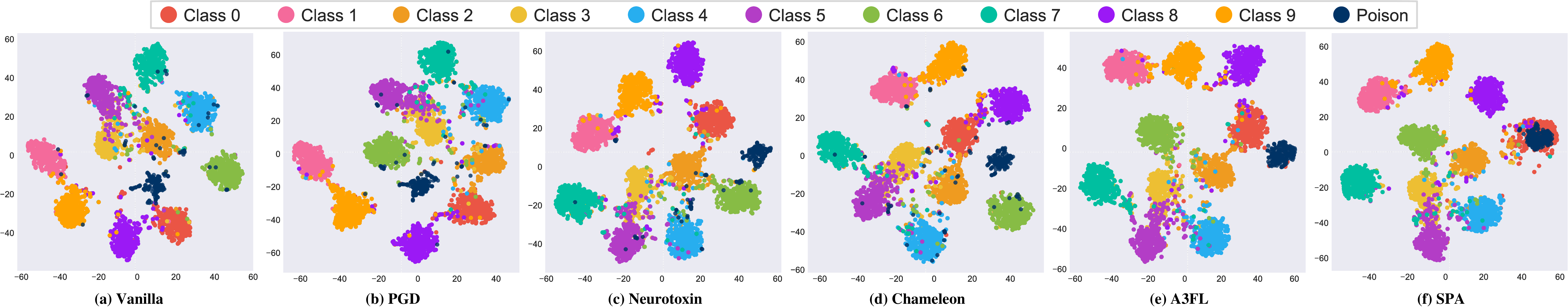}
	\caption{T-SNE visualization of \sysname compared with five attacks.} 
	\label{F13}
\end{figure*}

 We employ T-SNE visualization to further understand the feature distributions of different attacks. Specifically, we poisoned 10$\%$ of the test data in CIFAR-10 and fed them into backdoor models trained under different attack strategies, then visualized their output representations using T-SNE. As shown in Figure \ref{F13}, black points represent samples carrying triggers while other colors represent benign samples from different classes. The visualization reveals that feature representations of benign samples from the same class form distinct clusters, while poisoned samples form separate clusters. However, there exists a fundamental difference between our approach and traditional methods. Supervised-based methods establish a direct connection between target labels and backdoor features, resulting in backdoor clusters that are more compact and completely isolated from normal clusters, essentially creating an OOD backdoor attack. This significant divergence between backdoor and benign models makes such attacks detectable by existing defense mechanisms, particularly those based on OOD sample detection.

In contrast, our proposed \sysname method generates clusters that position themselves in proximity to the target class cluster. This strategic arrangement effectively misleads the model into perceiving backdoor trigger features as legitimate components of benign features. The visualization clearly demonstrates how our method blurs the boundary between backdoor and target class features in the feature space, creating a more natural integration rather than an obvious separation. This characteristic enables our attack to circumvent existing defense methods that rely on identifying anomalous feature clusters. The feature space visualization provides evidence for why SPA exhibits superior stealth and persistence compared to baseline methods. Traditional backdoor attacks create distinct feature representations that stand apart from normal distributions, effectively creating a separate path for classification that can be identified through careful analysis. Our approach, however, modifies the feature space more subtly by aligning backdoor features with legitimate target class features, making the distinction between poisoned and clean samples less obvious in the representation space.

\begin{figure}
	\centering
	\includegraphics[width=0.70\linewidth]{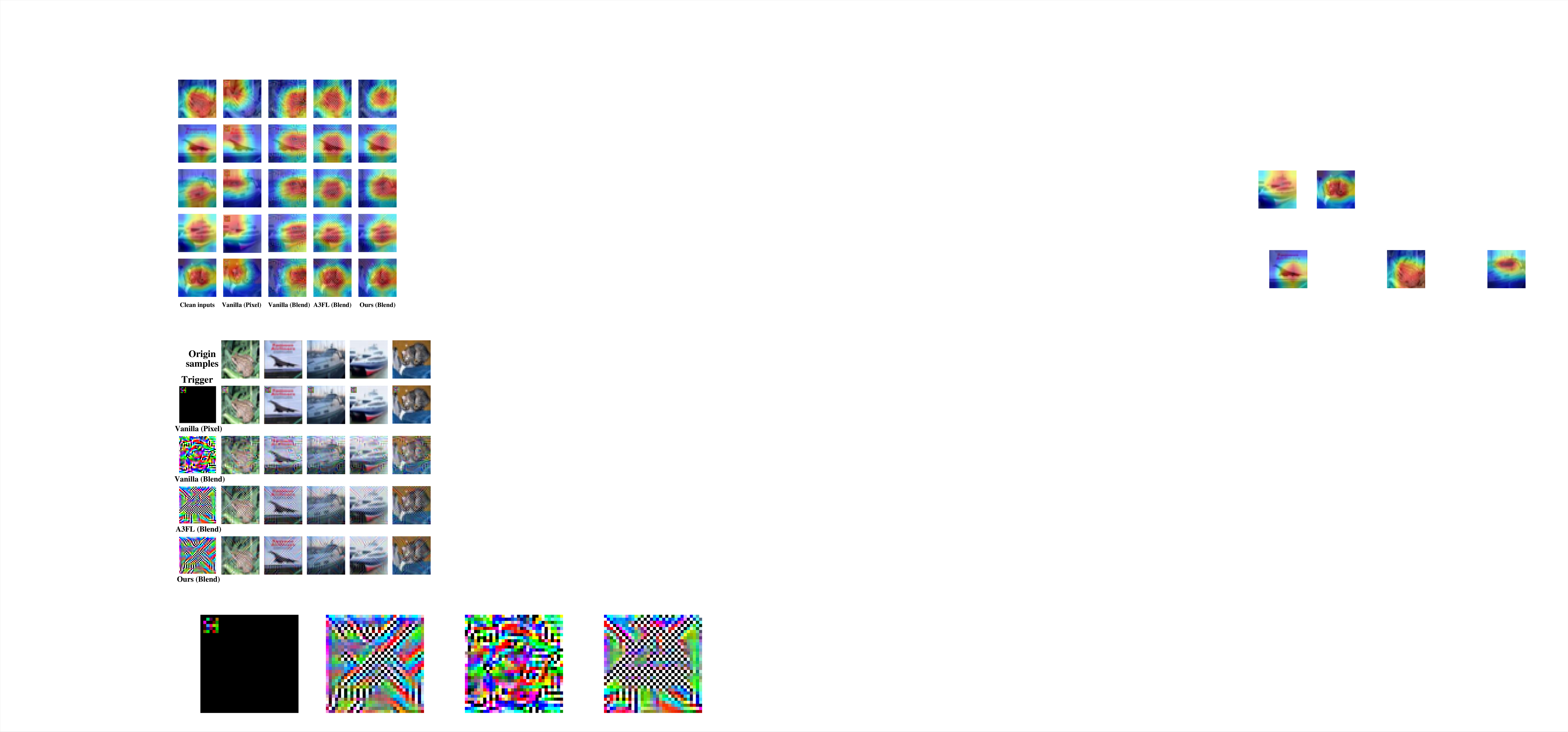}
	\caption{Visualization of different triggers and poisoned samples.}
	\label{F14}
	\vspace{-3mm}
\end{figure}

\begin{figure}
	\centering
	\includegraphics[width=0.65\linewidth]{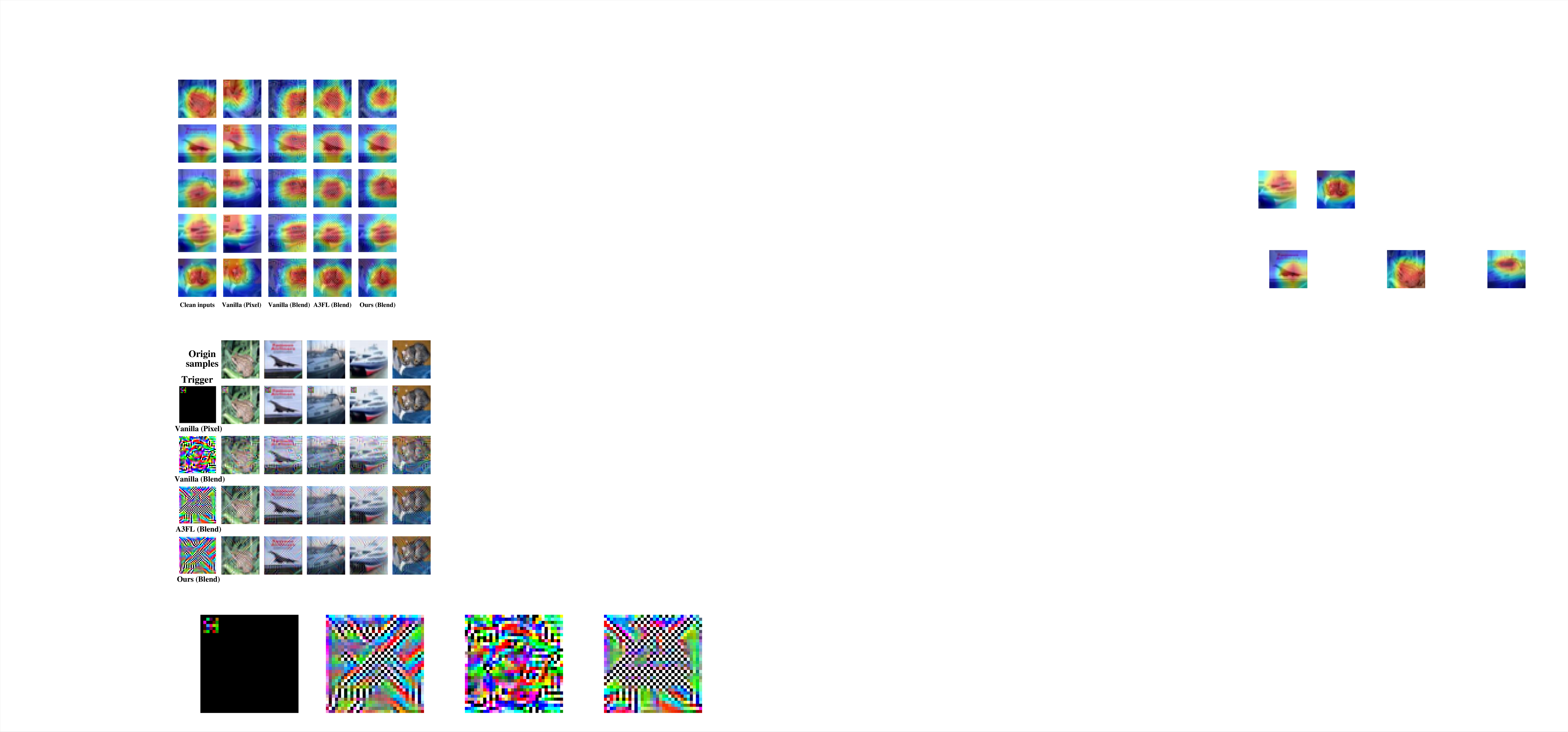}
	\caption{Grad-CAM visualization of different poisoned samples.}
	\label{F15}
	\vspace{-3mm}
\end{figure}

\begin{table}[t]
 \caption{\small Performance of \sysname with different triggers.}
  \vspace{1mm}
  \centering
  \footnotesize
  \renewcommand\tabcolsep{10pt}
  \renewcommand\arraystretch{1.2}
\begin{tabular}{ccccc}
\toprule
\multicolumn{2}{c}{\textbf{Type}}                                     & \textbf{Dataset} & \textbf{ACC} & \textbf{ASR} \\ \midrule
\multirow{6}{*}{\textbf{Fixed}}     & \multirow{3}{*}{\textbf{Pixel}} & CIFAR-10         & 90.09        & 53.97        \\
                                    &                                 & CIFAR-100        & 68.83        & 69.34        \\
                                    &                                 & GTSRB            & 95.11        & 80.46        \\ \cmidrule{2-5} 
                                    & \multirow{3}{*}{\textbf{Blend}} & CIFAR-10         & 89.22        & 68.05        \\
                                    &                                 & CIFAR-100        & 70.12        & 65.32        \\
                                    &                                 & GTSRB            & 94.42        & 90.24        \\ \midrule
\multirow{6}{*}{\textbf{Optimized}} & \multirow{3}{*}{\textbf{Pixel}} & CIFAR-10         & 92.58        & 97.9         \\
                                    &                                 & CIFAR-100        & 71.32        & 93.18        \\
                                    &                                 & GTSRB            & 96.25        & 99.49        \\ \cmidrule{2-5} 
                                    & \multirow{3}{*}{\textbf{Blend}} & CIFAR-10         & 92.55        & 99.91        \\
                                    &                                 & CIFAR-100        & 71.63        & 99.97        \\
                                    &                                 & GTSRB            & 96.79        & 99.57        \\ \bottomrule
\end{tabular}
  \label{T6}
 \vspace{-3mm}
\end{table}

\section{Different Trigger Types} 
\label{A-triiger_type}
In this section, we systematically evaluate the impact of different trigger patterns on attack performance. We first conduct a visual comparison of three distinct trigger types: one static pixel-based pattern and two optimized global blend triggers with a blending strength of 0.33. As demonstrated in Figure \ref{F14}, the blend triggers exhibit superior stealthiness compared to the pixel pattern, showing more natural integration with the original images. To further analyze their behavioral differences, we employ Grad-CAM visualizations to examine the model's attention distribution between clean and poisoned samples.  The results shown in Figure \ref{F15} reveal significant disparities in attention patterns across trigger types. For pixel-based triggers, the model's attention becomes heavily concentrated on the trigger region when processing poisoned samples, creating a stark contrast with its attention distribution on clean inputs. This conspicuous shift makes such attacks more detectable through attention-based analysis. In contrast, optimized blend triggers demonstrate more dispersed attention patterns, though still distinguishable from normal behavior. Our method achieves natural attention distribution, with poisoned samples showing minimal deviation from clean samples' attention maps. 

To quantify the performance implications of trigger types, we investigate the performance variation between optimized and non-optimized triggers. Experimental results shown in Table \ref{T6} confirm that dynamically optimized triggers consistently achieve higher ASR than their fixed counterparts, with blend triggers outperforming pixel patterns across all configurations. Based on our analysis, this performance difference stems from the inherent challenges of aligning OOD features, such as those induced by static pixel triggers, with the target feature distribution. The forceful alignment of disparate feature distributions can introduce distortions that compromise the target representation, as the alignment process is bidirectional and may inadvertently perturb the learned features. The superior performance of optimized triggers can be attributed to their ability to adapt dynamically during the training process. Rather than imposing predetermined patterns, optimized triggers evolve to identify and exploit the most effective pathways for feature alignment, resulting in more natural integration with the learned representation space. This adaptive quality enables the triggers to find minimal-resistance paths for feature manipulation, reducing the disruption to legitimate features while maintaining high attack effectiveness.



\end{document}